\documentclass[sigconf]{acmart}

\usepackage{acmart-taps}
\usepackage[utf8]{inputenc}
\usepackage{multirow}
\usepackage{makecell}
\usepackage{array}
\usepackage{cleveref}
\usepackage{subcaption}
\usepackage{pifont}
\usepackage{soul}
\usepackage{enumitem}
\usepackage{xspace}

\newcolumntype{C}[1]{>{\centering\arraybackslash}p{#1}}
\newcolumntype{L}[1]{>{\raggedright\arraybackslash}p{#1}}
\newcolumntype{Y}{>{\centering\arraybackslash}X}

\newcounter{findingnum}

\usepackage{pifont}  
\newcommand{\mainfinding}[2]{%
  \item[\hspace{1em}\ding{\numexpr201+#1\relax}] #2%
}

\definecolor{insufforange}{RGB}{220, 110, 0}   
\definecolor{suffgreen}{RGB}{40, 140, 40}      
\newcommand{\insufficient}{\textcolor{insufforange}{\textit{Insufficient}}\xspace}
\newcommand{\sufficient}{\textcolor{suffgreen}{\textit{Sufficient}}\xspace}

\setul{1pt}{1.25pt}  

\definecolor{earlyblue}{RGB}{200, 230, 255}     
\definecolor{contblue}{RGB}{150, 200, 255}      
\definecolor{lateblue}{RGB}{0, 70, 180}         
\definecolor{noblue}{RGB}{0, 40, 140}           

\aptLtoX{}{\newcommand{\earlyai}{{\setulcolor{earlyblue}\ul{\textit{Early}}}\xspace}
\newcommand{\continuousai}{{\setulcolor{contblue}\ul{\textit{Continuous}}}\xspace}
\newcommand{\lateai}{{\setulcolor{lateblue}\ul{\textit{Late}}}\xspace}
\newcommand{\noai}{{\setulcolor{noblue}\ul{\textit{No}}}\xspace}}

\AtBeginDocument{%
  }

\setcopyright{acmlicensed}

\copyrightyear{2026}
\acmYear{2026}
\setcopyright{cc}
\setcctype{by}
\acmConference[CHI '26]{Proceedings of the 2026 CHI Conference on Human Factors in Computing Systems}{April 13--17, 2026}{Barcelona, Spain}
\acmBooktitle{Proceedings of the 2026 CHI Conference on Human Factors in Computing Systems (CHI '26), April 13--17, 2026, Barcelona, Spain}
\acmPrice{}
\acmDOI{10.1145/3772318.3791796}
\acmISBN{979-8-4007-2278-3/2026/04}

\newcommand{\edit}[1]{\textcolor{black}{#1}}

\begin{document}

\title[Effects of LLM Use on Critical Thinking]{Investigating the Effects of LLM Use on Critical Thinking\\Under Time Constraints: Access Timing and Time Availability}


\author{Jiayin Zhi}
\orcid{0009-0006-9290-7356}
\email{jzhi@uchicago.edu}
\affiliation{%
 \institution{University of Chicago}
 \city{Chicago}
 \state{Illinois}
 \country{USA}}

\author{Harsh Kumar}
\orcid{0000-0003-2878-3986}
\email{harsh@cs.toronto.edu}
\affiliation{%
 \institution{University of Toronto}
 \city{Toronto}
 \state{Ontario}
 \country{Canada}}

\author{Mina Lee}
\orcid{0000-0002-0428-4720}
\email{mnlee@uchicago.edu}
\affiliation{%
 \institution{University of Chicago}
 \city{Chicago}
 \state{Illinois}
 \country{USA}}

\renewcommand{\shortauthors}{Zhi et al.}

\begin{abstract}

The impact of large language models (LLMs) on critical thinking has provoked growing attention, yet this impact on actual performance may not be uniformly negative or positive. Particularly, the role of time---the temporal context under which an LLM is provided---remains overlooked. In a between-subjects experiment (n=393), we examined two types of time constraints for a critical thinking task requiring participants to make a reasoned decision for a real-world scenario based on diverse documents: (1) LLM access timing---an LLM available only at the beginning (early), throughout (continuous), near the end (late), or not at all (no LLM), and (2) time availability---insufficient or sufficient time for the task. We found a temporal reversal: LLM access from the start (early, continuous) improved performance under time pressure but impaired it with sufficient time, whereas beginning the task independently (late, no LLM) showed the opposite pattern. These findings demonstrate that time constraints fundamentally shape whether an LLM augments or undermines critical thinking, making time a central consideration when designing LLM support and evaluating human-AI collaboration in cognitive tasks. 

\end{abstract}





\begin{teaserfigure}
  \centering
  \includegraphics[width=0.95\textwidth]{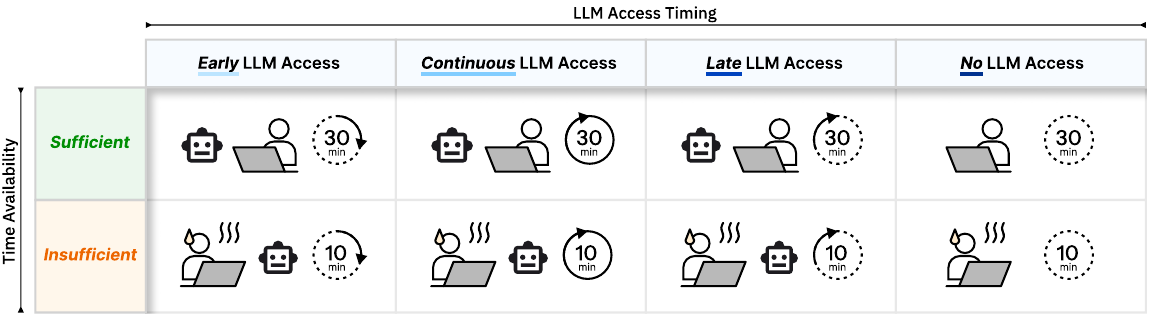}
  \caption{This $4 \times 2$ experimental framework tests the effects of LLM use on critical thinking under two types of time constraints: access timing (x-axis) and time availability (y-axis). Participants, working under time pressure (\insufficient) or not (\sufficient), were randomized to receive LLM access only at the beginning of the task (\earlyai LLM access), throughout (\continuousai LLM access), near the end (\lateai LLM access), or not at all (\noai LLM access).
  }
  \label{fig:time_constraint_framework}
\end{teaserfigure}

\maketitle

\enlargethispage{20pt}
\section{Introduction}

Critical thinking---the ability to reason through diverse and sometimes conflicting information by analysis, evaluation, and synthesis to reach reasoned decisions---is fundamental for our work and life \cite{shavelson2019assessment,braun2020performance,krathwohl2002revision}. 
In online information environments, whether navigating conflicting news, or making judgments based on scattered documents, people always face situations that require critical thinking and apply the reasoned conclusions to various purposes (e.g., solve a problem, decide on an action, or answer a given question), which involves distilling argumentative information from relevant sources, analyzing source credibility, avoiding bias, and communicating reasoning succinctly \cite{gerlich2025ai,shavelson2019assessment,braun2020performance,krathwohl2002revision}.


As generative AI (GenAI), especially large language models (LLMs), have been increasingly integrated into information technologies, critical thinking is highly exposed to them. In fact, critical thinking is the most frequently required capability in tasks that users bring to popular LLM-powered chatbots \cite{handa2025economic}. 
\edit{Recent user research provides valuable initial insights on the connection between AI use and critical thinking.} 
\edit{A survey collected examples of how knowledge workers enact critical thinking when using GenAI for different work tasks \cite{lee2025impact}. A mixed-methods study investigating how general users perceive their critical thinking abilities found that using AI tools can reduce their cognitive effort \cite{gerlich2025ai}.} For example, ready-made AI summaries eliminate the need to comprehend sources or evaluate evidence \cite{gerlich2025ai}. 
\edit{However, these methods fall short in establishing causal relationships and objectively measuring performance, leaving the actual performance consequences unclear. Beyond user research and self-reports, experiments using tasks that assess critical thinking performance are needed to determine how the impact of LLMs actually manifests.}

Moreover, the impact of LLMs on critical thinking may not be uniformly negative or positive, but depend on the temporal context in which the LLM is provided. Two types of time constraints emerge as particularly crucial for human-AI collaboration in cognitive tasks requiring extended reasoning. First, the timing of LLM access. Recent studies show the benefits of having \edit{LLM} access in the late time period for solving math problems \cite{kumar2025math} and creative writing \cite{qin2025timing}. 
However, for critical thinking, there are also reasons to hope that early access to an LLM could be beneficial. It could serve as initial scaffolding and handle preliminary groundwork---such as gathering available information and grasping basic concepts---potentially freeing cognitive resources for deep deliberation \cite{krathwohl2002revision,singh2025protecting, imundo2024expert,lee2025impact}. 
Second, time availability. In practice, people often face time pressure for the tasks they do, whether from explicit deadlines or implicit expectations \cite{zuzanek2004work,baethge2018matter,kelly1988entrainment,lee2025impact,hoge2009work}. 
Research suggests that time pressure can shape cognitive performance, shifting people from deliberative reasoning to heuristic processing \cite{kocher2006time,moore2012timepressure,kelly1988entrainment}.
These two temporal dimensions inherently interact in reality, yet previous research has examined a subset of conditions within each temporal dimension in isolation across disparate domains. This fragmented picture leaves us with critical gaps in understanding how time constraints shape LLMs' impact on critical thinking under realistic conditions where both temporal dimensions operate simultaneously.


In this work, we investigated \textbf{how LLMs influence participants' critical thinking task performance under two types of time constraints: (1) LLM access timing and (2) time availability.} We tested four LLM access timings: \earlyai, \continuousai, \lateai, and \noai LLM access, where an LLM was available only at the beginning of the task, throughout, near the end, or not at all, respectively. For time availability, participants had either \insufficient or \sufficient time for the task. Based on these time constraints, we asked:




\begin{description}[leftmargin=2.5em, labelwidth=1em, labelsep=0em, style=sameline]
  \item[\textbf{RQ1.}] How does LLM access timing affect critical thinking task performance under different time availability?
  \item[\textbf{RQ2.}] How does time availability affect critical thinking task performance when having the same LLM access timing?
\end{description}

To answer these questions, we conducted a preregistered\footnote{https://aspredicted.org/q6yh-mrqg.pdf} $4 \times 2$ between-subjects experiment ($n=393$). 
\edit{We employed the iPAL (International Performance Assessment of Learning) framework \cite{shavelson2019assessment,braun2020performance,krathwohl2002revision} developed for assessing the authentic performance of critical thinking.}
\edit{A critical thinking performance assessment task} \cite{braun2020performance,EbrightJonesCortina2025} ask participants to make a reasoned decision for a real-world scenario in civic life based on a set of documents of varying characteristics.
Participants need to navigate these documents, analyze, and draw inferences from evidence, evaluate the sources, and synthesize conflicting perspectives into a reasoned written decision---capturing the non-linear, reciprocal mental processes that characterize critical thinking.
Under \insufficient or \sufficient time, participants were randomly assigned to one of the four \edit{LLM} access timings, and completed the task. Performance was primarily evaluated based on the Essay written as part of the task. Then, we additionally measured participants' performance of Recall (remembering provided documents), Evaluation (evaluating the characteristics of source documents), and Comprehension (understanding documents by evidence-based inference), to capture cognitive activities that could influence critical thinking. After the task, participants completed a critical thinking self-assessment \cite{payan2022development, tanprasert2024debate}.

Overall, our results revealed that the effect of LLM use on critical thinking task performance fundamentally depends on LLM access timing and time availability (Section \ref{main_results}). For \textbf{RQ1}, under \insufficient time, having \edit{the LLM} from the start (\earlyai and \continuousai LLM access) increased Essay performance of critical thinking, compared to working independently first (\lateai and \noai LLM access). Yet under \sufficient time, this pattern dramatically reversed---those who worked independently first (\lateai and \noai LLM access) showed better Essay performance. Moreover, under \sufficient time, having \edit{the LLM} access from the start (\earlyai and \continuousai LLM access) impaired Recall performance, suggesting it might prevent internalization of the source documents. 
For \textbf{RQ2}, having \sufficient time substantially improved Essay and Recall performance for participants who worked independently first (\lateai and \noai LLM access), but provided minimal benefits to those who had \edit{the LLM} from the start (\earlyai and \continuousai LLM access). \edit{Meanwhile, self-assessment showed minimal variation across conditions, suggesting its limitations in detecting how LLMs affect critical thinking under time constraints.} Further exploration of participants' interaction with their assigned LLM access timing offered insights into the mechanisms underlying the results (Section \ref{behavior_engagement}). Our interface code, data, and analysis are available at the project website.\footnote{\url{https://criticalthinking-ai.app}}


Our work demonstrates the importance of considering time constraints when discussing the benefits or harms of using LLMs for critical thinking. Concretely, we make the following contributions: 
\begin{itemize}
  \item{Rich empirical findings on the impact of LLMs on critical thinking, moving beyond simple comparisons of using LLMs or not, varied by LLM access timing and time availability;}
  \item{Recommendations for designing LLM support for tasks requiring critical thinking;}
  \item{Implications for human-AI collaboration research, emphasizing the need to consider time constraints.}
\end{itemize}


\section{Related Work}

\subsection{Critical Thinking and Its Assessment}

Critical thinking---the ability to analyze, synthesize, and evaluate information to reach reasoned decisions---involves higher-order cognitive processes that play an important role in learning \cite{liu2014assessing,erwin2003assessment,krathwohl2002revision,shavelson2019assessment,braun2020performance}. In Bloom's taxonomy, these higher-order processes are related to lower-order activities---remembering, understanding, and applying---which require individuals to internalize and comprehend information as the foundation \cite{krathwohl2002revision,armstrong2010bloom,oakley2025memory}. Education and psychology researchers have long studied how to measure critical thinking, primarily using three approaches: standardized tests, self-report measures, and performance assessments.

Standardized tests present participants with short scenarios followed by questions. Multiple-choice formats like the Watson-Glaser Critical Thinking Appraisal present short passages followed by multiple-choice questions about drawing inferences, recognizing assumptions, or identifying logical flaws \cite{watson1980wgcta,sternod2016wgcta,facione1990cctst,ennis1985cornell}. Hybrid formats like the Halpern Critical Thinking Assessment combine multiple-choice and short constructed-response questions \cite{butler2012halpern}. For instance, a participant might read a brief scenario about a new medication and be asked, \textit{``What is the flaw in this reasoning?''}---explicitly directing attention to a predetermined issue. However, these assessments have certain limitations: they do not capture authentic critical thinking because participants are explicitly prompted rather than independently identifying problems, they emphasize deterministic logic, and they remain divorced from realistic information environments where critical thinking involves analyzing, evaluating, and synthesizing across multiple sources \cite{liu2014assessing,shavelson2019assessment,EbrightJonesCortina2025}.


Self-report measures ask participants to rate their own critical thinking. For instance, the Critical Thinking Self-Assessment Scale asks participants to rate statements, such as \textit{``I examine the logical strength of the underlying reason in an argument''} \cite{payan2022development}.
Prior work has shown the risks of self-reports that participants tend to overestimate their abilities \cite{cole2010accuracy}, and can show poor metacognitive accuracy when assessing their own performance \cite{double2025survey}.

Performance assessments address these limitations by placing participants in a realistic decision-making scenarios with a constructed set of documents and ask them to write argumentative essays explaining their reasoning. 
The iPAL (International Performance Assessment of Learning) framework represents the most recent approach for performance assessments \cite{shavelson2019assessment,braun2020performance}, addressing the challenges of subjective scoring in earlier frameworks \cite{ennis1985ewctet,klein2007cla,cae2006cla}. 
It has been widely validated across diverse populations in education and psychology research, showing that it captures multiple aspects of critical thinking with good inter-rater reliability \cite{hahnel2019validating, zlatkin2021advances,nagel2020performance,EbrightJonesCortina2025,hyytinen2019developing,shavelson2019assessment,ipal_publications}. Known-groups validation further supports construct validity: participants receiving critical thinking training consistently outperform the control group \cite{EbrightJonesCortina2025, nagel2020performance}. It has been applied in empirical investigations from everyday contexts to higher education \cite{munchow2019ability, hyytinen2024self,molerov2020assessing,anghel2021college} and professional domains, such as administration \cite{hyytinen2024self,atan2024empirical}, education \cite{jeschke2019performance,garcia2021qualitative}, and economics \cite{zlatkin2021advances}.

This framework demonstrates strong ecological validity by simulating high-density information environments that mirror real-world civic participation, 
which allows researchers to observe participants' critical thinking performance, not just whether they can identify a flaw when prompted. Without realizing their critical thinking is being assessed, participants must locate, analyze, evaluate, and synthesize conflicting source information to reach and communicate reasoned decisions \cite{braun2020performance,ennis1985ewctet,klein2007cla,cae2006cla}. Thus, it is well-suited to capture the authentic, non-linear process of critical thinking across multiple facets. In addition, because all argumentative information exists within this constructed information environment, scoring becomes objective, as it is primarily based on the quantity and quality of arguments participants synthesize and present, rather than making holistic judgments about rhetorical or writing skills \cite{shavelson2019assessment, braun2020performance}. This approach also reduces construct-irrelevant variance by minimizing confounds from prior background knowledge \cite{braun2020performance, EbrightJonesCortina2025}---the assessment measures how participants reason with the given evidence, not what they already know.

\subsection{LLMs' Influence on Critical Thinking}


As LLMs become ubiquitous, there is growing attention to their impact on critical thinking. Recent user research provides valuable initial insights on this connection, grounding the discussion in user-reported experiences across personal and professional contexts. For instance, \citet{gerlich2025ai} found that general users who used AI tools more frequently reported reduced critical thinking abilities. \citet{lee2025impact} examined how knowledge workers enact critical thinking while using GenAI, finding that higher confidence in GenAI was associated with reduced enaction of critical thinking, while higher self-confidence was associated with greater enaction. However, these early studies rely on how users perceive their own critical thinking abilities and behaviors, rather than objective performance measures. Actual behavior may diverge from what users perceive---especially for complex reasoning where metacognitive blind spots are common \cite{double2025survey,cole2010accuracy}.



What remains less understood is how LLM use affects actual performance on tasks requiring critical thinking, as experimental evidence on this is still limited. The experimental study closest to our work is by \citet{melumad2025experimental}, which offers valuable insights into how LLMs affect information seeking and online reasoning in everyday contexts. Their study compared participants using an LLM or a traditional search engine for writing gardening advice. 
They found that participants using the LLM reported reduced depth of learning, spent less time on the task, and produced sparser advice. However, the task was relatively low-stake and did not necessarily require evaluating conflicting sources or synthesizing competing arguments. Additionally, learning was measured as a unitary construct through self-reported ratings, and without a controlled information environment, performance may be influenced by participants' prior background knowledge. 

Our work addresses these gaps in prior studies to investigate LLMs' impact on critical thinking. Using the performance assessment framework, we objectively assess critical thinking through a controlled information environment where participants need to process and reason through novel documents of varying relevance, trustworthiness, and stances to resolve a civic decision-making scenario, minimizing confounds from background knowledge.

\begin{table*}[htbp]
\centering
\small
\renewcommand{\arraystretch}{1}
\begin{tabular}{@{}p{2.5cm}|l|l|l|l@{}}
\toprule
& \textbf{Early LLM access} & \textbf{Continuous LLM access} & \textbf{Late LLM access} & \textbf{No LLM access} \\
\midrule
\textbf{Insufficient time} & 
& 
& 
& 
\begin{tabular}[t]{@{}l@{}}[Reasoning] \textit{Insufficient} <\\ \textit{Sufficient} \cite{gonthier2023should}\end{tabular} \\
\midrule
\textbf{Sufficient time} &
\begin{tabular}[t]{@{}l@{}}\mbox{}\\[1.7em] [Math education] \textit{Early} <\\ \textit{Late} \cite{kumar2025math}\end{tabular}
& 
\begin{tabular}[t]{@{}l@{}}[Creative writing] \textit{Continuous} <\\ \textit{Late} \cite{qin2025timing}\end{tabular}
& 
\begin{tabular}[t]{@{}l@{}}[Creative writing] \textit{Continuous} <\\ \textit{Late} \cite{qin2025timing}\\[0.5em] [Math education] \textit{Early} <\\ \textit{Late} \cite{kumar2025math}\end{tabular}
& 
\begin{tabular}[t]{@{}l@{}}[Reasoning] \textit{Insufficient} <\\ \textit{Sufficient} \cite{gonthier2023should}\end{tabular} \\
\bottomrule
\end{tabular}
\caption{Summary of prior work with time constraints in human-AI collaboration for learning and thinking outcomes. Each cell reports the study domain and the direction of the observed effect in the format: [domain] effect direction. Empty cells indicate conditions where we identified no recent empirical studies. The table includes only studies that explicitly manipulate time constraints; most prior work comparing LLM use with no LLM without time constraints is omitted. 
}
\label{tab:related-work-time-constraints}
\end{table*}


\subsection{Time Constraints in Human-AI Collaboration}


Drawing from previous literature (Table~\ref{tab:related-work-time-constraints}), we identify two types of time constraints that can shape human-AI collaboration for thinking and learning: the timing of LLM access and the time availability for task completion.

(1) LLM access timing. Recent work suggests the benefits of late LLM access for thinking and learning outcomes. In math education, \citet{kumar2025math} found that LLM explanation was most beneficial for test performance when students attempted problems independently before receiving it during practice. In creative writing, \citet{qin2025timing} demonstrated that using LLMs continuously throughout the task reduced the originality of ideas compared to using LLMs later in the second half of the given task time. However, there are reasons to hope that early LLM access may also offer benefits, particularly for critical thinking. LLMs could serve as initial scaffolding, handling lower-order cognitive activities such as information retrieval and initial comprehension that happens more often in early time period of a task \cite{imundo2024expert,singh2025protecting}, potentially freeing cognitive resources for higher-order thinking activities \cite{krathwohl2002revision}. The optimal timing of LLM access for critical thinking remains an open empirical question.

(2) Time availability. Research in psychology, management, and organizational behavior has long established that time pressure can shape human cognition \cite{kelly1988entrainment,moore2012timepressure,zhao2022temporal,kocher2006time,depaola2016timepressure,ordonez1997timepressure,karau1992timescarcity}. For example, \citet{gonthier2023should} found that under time pressure, people have impaired performance in reasoning tasks, showing limited capacity for deliberation and reliance on heuristic processing. In the HCI literature, \citet{cao2023timepressure} examined time pressure on participants' agreement with AI recommendations, and \citet{swaroop2024accuracy} examined both time pressure and ordering of AI recommendations and human decisions. Yet these studies focus on whether participants agree with AI recommendations or not in discrete-choice tasks---a different mechanism from open-ended thinking and learning tasks that require extended reasoning.

In practice, these two temporal dimensions inherently interact: access timing operates within the bounds of time availability. Yet they have been studied in isolation with many comparisons unexplored (Table~\ref{tab:related-work-time-constraints}). In our study, we examine the effects of LLM access timing under different time availability conditions, and vice versa, to establish a systematic understanding of how time constraints shape human-AI collaboration in critical thinking.

\section{Method}



\subsection{Study Design}

\subsubsection{Overview}
\begin{figure*}[htbp]
    \centering
    \begin{subfigure}[t]{0.48\textwidth}
        \centering
        \includegraphics[width=\textwidth]{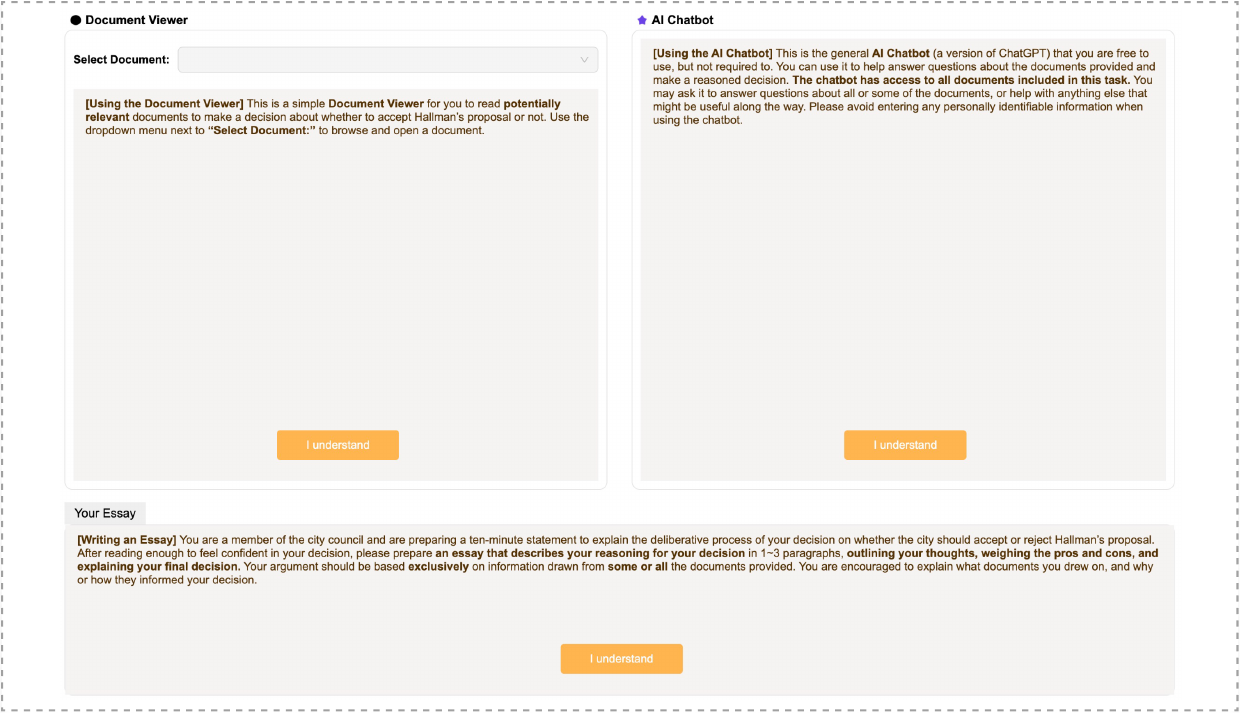}
        \caption{Instructions of the document viewer (left), LLM-powered chatbot with access to all the documents (right), and essay textbox (bottom). The timer begins after participants click all ``I understand'' buttons.}
        \label{fig:ai_instruction}
    \end{subfigure}\hfill%
    \begin{subfigure}[t]{0.48\textwidth}
        \centering
        \includegraphics[width=\textwidth]{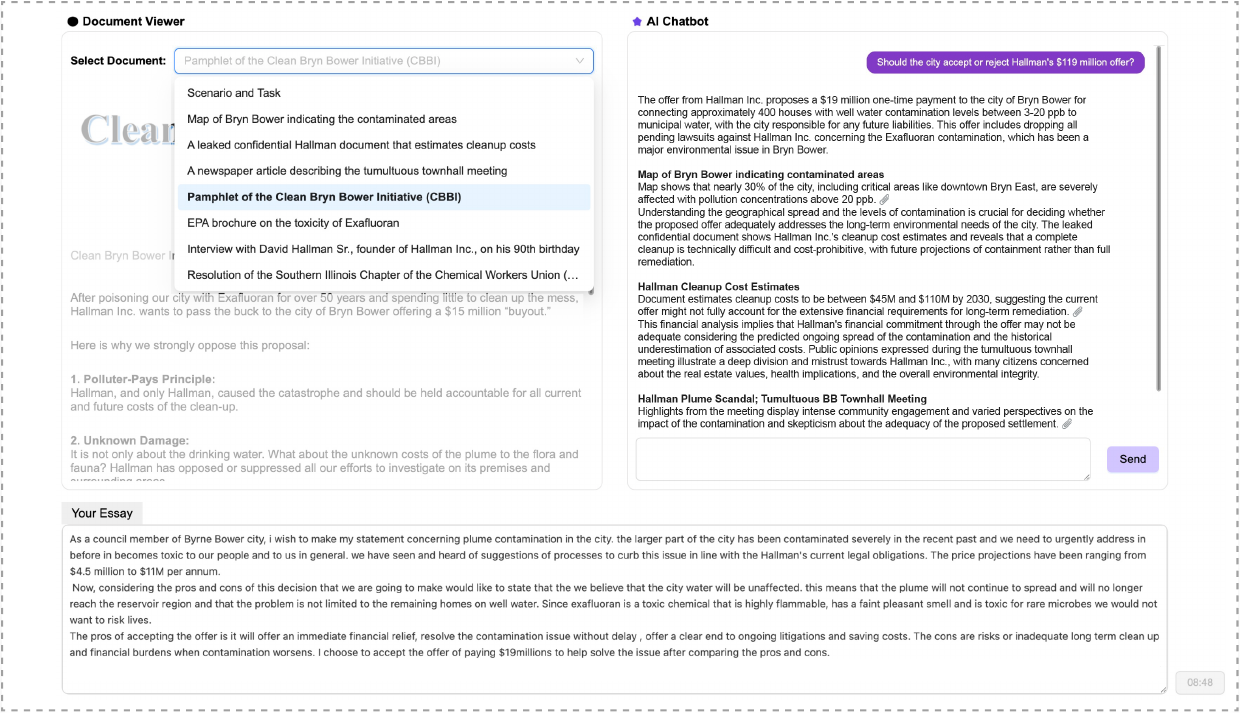}
        \caption{The interface controls the LLM access and time availability. The timer at the bottom right shows the remaining task time. Participants are automatically taken to the next page when time runs out.}
        \label{fig:ai_inuse}
    \end{subfigure}
    
    \caption{Interface for the main task in the six conditions having LLM access (\earlyai, \continuousai, \lateai LLM access under \insufficient and \sufficient time availability): (a) instructions and (b) an example of the active interface during the task. Participants in two conditions without LLM access (\noai LLM access under \insufficient and \sufficient time availability) only have the document viewer and the essay textbox.}
    \label{fig:interface}
\end{figure*}

We designed a 4×2 between-subjects experiment to investigate how LLM access timing influences participants' critical thinking task performance under different time constraints. We manipulated LLM access timing across four conditions (\earlyai, \lateai, \continuousai, and \noai LLM access) and time availability across two conditions (\insufficient and \sufficient). 
We employed a task from the critical thinking performance assessment framework \cite{ebright2024we,EbrightJonesCortina2025,braun2020performance,shavelson2019assessment}, which asks participants to write an essay justifying their decision on a civic decision-making scenario based on a curated set of documents of varying relevance, trustworthiness, and stances, regardless of specialized background knowledge. Using a custom web interface (Figure \ref{fig:interface}), each participant completed the task with their randomly assigned access timing to an LLM-powered chatbot, under \insufficient or \sufficient time availability.

\subsubsection{Experimental conditions}\label{conditions}

Our experimental design manipulated two types of time constraints: LLM access timing and time availability. Under different time availability (\insufficient and \sufficient), participants were randomly assigned to one of four LLM access timings (\earlyai, \continuousai, \lateai, and \noai LLM access). This factorial design (Figure \ref{fig:time_constraint_framework}) allowed us to examine both the main and interaction effects of the two types of time constraints. \edit{Before publishing the main study, we conducted four pilot studies to iterate on the time constraints. Pilot 1 ($n=60$) randomly assigned participants to have LLM access throughout the task or no LLM access, without time constraints, testing the study interface and procedure. Pilots 2 ($n=40$), 3 ($n=80$), and 4 ($n=160$) progressively scaled up on the full factorial design, with half of participants under \insufficient and half under \sufficient time availability, randomly assigned to one of four LLM access timings.}

For LLM access timing, we created four levels varying when participants could access the LLM during the task: \earlyai (LLM accessible in the first third of task time), \continuousai (LLM accessible throughout), \lateai (LLM accessible in the final third of task time), and \noai LLM access. This division was informed by Pilot 1, which showed participants without time constraints typically allocated approximately one-third of their cursor activity to the LLM-powered chatbot area. 
Pilots 2 to 4 confirmed that this manipulation was effective. 

For time availability, we established two levels: \insufficient (10 minutes) and \sufficient (30 minutes). Following practices by \citet{karau1992timescarcity}, we 
determined what time induces pressure and what time permits unhurried completion based on pilot studies. Pilot 1 participants without any time constraints averaged 20 minutes to complete the task. Further pilot testing (Pilots 2 to 4) confirmed that participants experienced 10 minutes as \insufficient (inducing time pressure) and 30 minutes as \sufficient (allowing thorough deliberation).






\subsubsection{Performance assessment task}
\label{sec:task}

We employed a task under the critical thinking performance assessment framework \cite{braun2020performance,shavelson2019assessment}, specifically the water contamination scenario \cite{ebright2024we,EbrightJonesCortina2025} (see cited papers for full task documents).
The task presents a real-world scenario where participants act as city council members deciding whether to accept a company's proposal to address the water contamination it caused, in exchange for dropping lawsuits and future liabilities. Participants are asked to prepare an essay explaining the deliberative process behind their decision for an upcoming council meeting, based exclusively on information drawn from some or all the provided documents. As shown in Table \ref{tab:document_overview}, the set of documents contain varying degrees of relevance (to the decision), trustworthiness (of the sources), and stance (pro, con, or neutral toward the proposal), representing diverse information sources including technical reports, newspaper articles, advocacy group pamphlets, and government agency brochures.

\begin{table*}[htbp]
\centering
\small
\renewcommand{\arraystretch}{1}
\setlength{\tabcolsep}{1pt}
\begin{tabular*}{\textwidth}{@{\extracolsep{\fill}}c p{3.5cm} p{7.5cm} l l l@{}}
\toprule
\textbf{} & \textbf{Document title} & \textbf{Summary of main idea} & \textbf{Relevance} & \textbf{Trustworthiness} & \textbf{Stance} \\
\midrule
1 & Map of Bryn Bower indicating contaminated areas & Showcases the range and extent of water contamination in Bryn Bower. & Relevant & Trustworthy & Neutral \\
\midrule
2 & A leaked Hallman document estimating cleanup costs & Provides estimates of cleanup costs for the water contamination, describing remediation expenses. & Relevant & Trustworthy & Pro/Neutral \\
\midrule
3 & A newspaper article describing tumultuous townhall meeting & Details a heated townhall meeting over Hallman's proposal, revealing divisions among Bryn Bower residents. & Mixed & Untrustworthy & Neutral/Con \\
\midrule
4 & Pamphlet of the Clean Bryn Bower Initiative & The Clean Bryn Bower Initiative opposes Hallman's proposal, arguing Hallman should fully bear cleanup responsibilities and criticizing its past environmental negligence. & Relevant & Untrustworthy & Con \\
\midrule
5 & EPA brochure on toxicity of Exafluoran & Outlines Exafluoran's potential health risks at various concentration levels, acknowledging that risks at low levels are unknown. & Relevant & Trustworthy & Neutral \\
\midrule
6 & Interview with David Hallman Sr., founder of Hallman Inc., on his 90th birthday & David Hallman discusses his reflections on environmental issues, and highlights the company's industrial contributions. & Irrelevant & Mixed & Pro \\
\midrule
7 & Resolution of the Southern Illinois Chapter of the Chemical Workers Union & The Chemical Workers Union supports Hallman's proposal to manage financial uncertainty from mandatory cleanup efforts, while acknowledging financial constraints. & Relevant & Trustworthy & Pro \\
\bottomrule
\end{tabular*}
\caption{Overview of documents provided in critical thinking performance assessment task \cite{ebright2024we,EbrightJonesCortina2025,braun2020performance}: document title, summary of main idea, and characteristics to be evaluated for critical thinking, including relevance to the decision, trustworthiness of the sources, and stance toward the proposal. The documents appeared in random order during the experiment.}
\label{tab:document_overview}
\end{table*}

\subsubsection{Study procedure}\label{study_procedure}


The study consisted of three phases: a pre-task survey, the main critical thinking performance assessment task, and a post-task assessment (see Supplementary Materials for all questions). The pre-task survey collected demographic information, frequency of use and attitudes toward LLM applications, and confidence in both critical thinking abilities and LLM capabilities.

\edit{For the main task, participants were first introduced to the task and the interface, given their assigned condition. Participants in conditions with LLM access were told they had access to a general AI chatbot (a version of ChatGPT) that could answer questions about all the documents provided, along with the timing of its availability. Participants were instructed not to use other tools and were informed they would receive an alert if they left the study interface. Participants then read the task scenario and instruction to prepare an essay reasoning about their decision, before proceeding to the main task interface where they performed the task (Section~\ref{sec:task}) until time expired.}

\begin{figure}[htbp]
    \centering
    \begin{subfigure}[t]{0.48\textwidth}
        \centering
        \includegraphics[width=\textwidth]{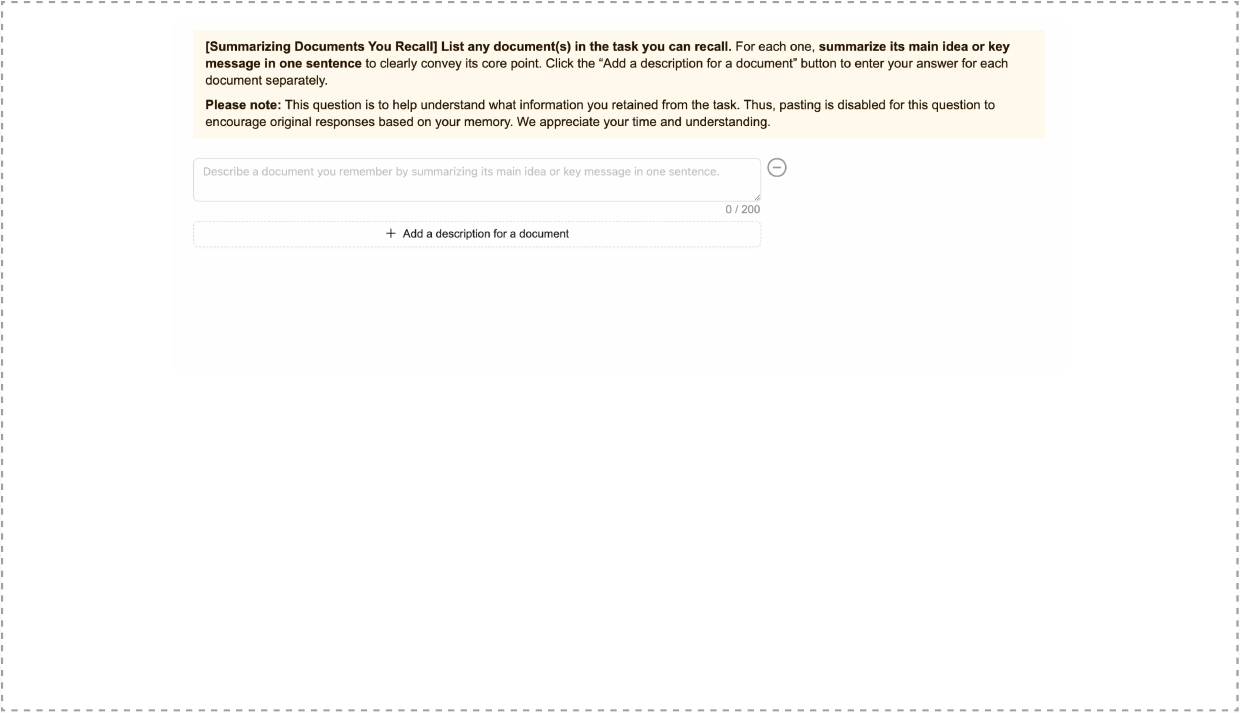}
        \caption{Participants can add fields to summarize the main ideas of documents by free recall. Pasting is disabled.}
        \label{fig:recall}
    \end{subfigure}\hfill%
    \vspace{0.5em}

    \begin{subfigure}[t]{0.48\textwidth}
        \centering
        \includegraphics[width=\textwidth]{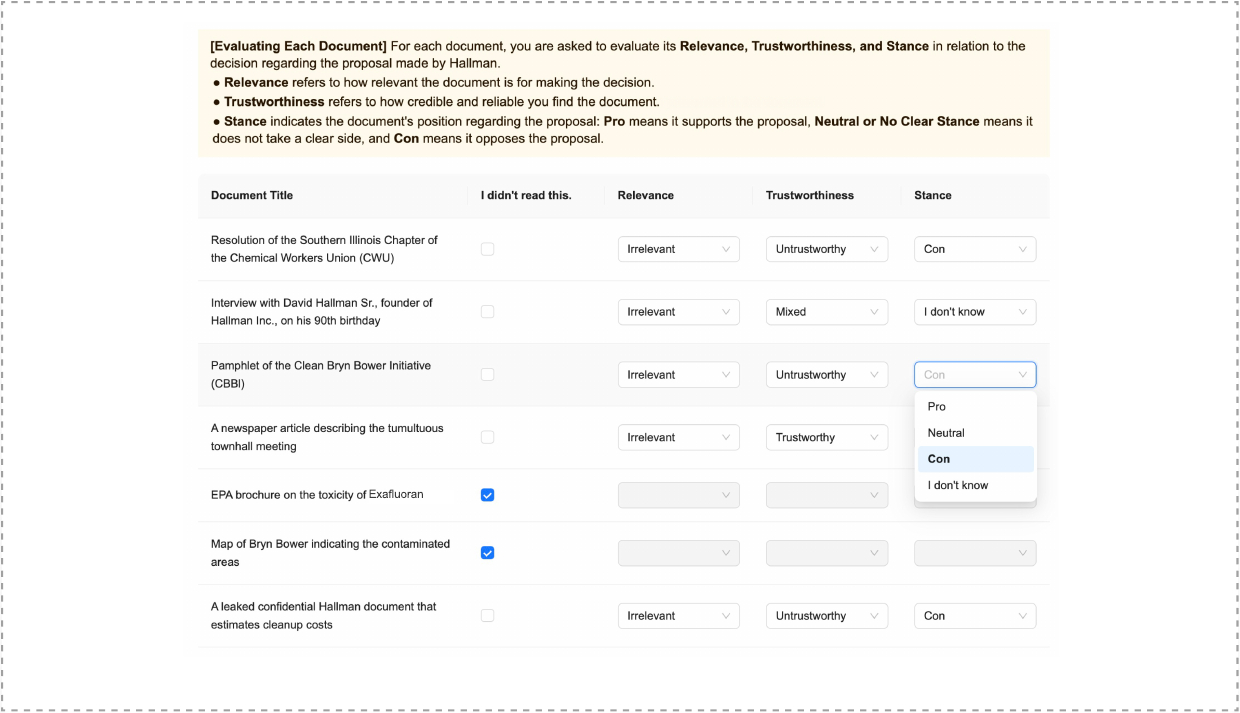}
        \caption{Participants rate the relevance, trustworthiness, and stance of each document with title provided. They can select ``I didn't read this'' for a document.}
        \label{fig:evaluation}
    \end{subfigure}
    
    \caption{Interface for assessing (a) Recall and (b) Evaluation performance. 
    }
    \label{fig:post_task_interface}
\end{figure}

\edit{The post-task assessment consisted of several components. First, participants explained how they approached the task in an open-ended response: those with LLM access were asked about how they used both their time and LLM access; those with \noai LLM access were asked about how they used their time. Next, they completed assessments for Recall, Evaluation, and Comprehension performance. For Recall, participants summarized each document's main idea through free recall without cues (Figure \ref{fig:recall}). For Evaluation, they rated each document's relevance (relevant, mixed, irrelevant, or I don't know), trustworthiness (trustworthy, mixed, untrustworthy, or I don't know), and stance (pro, con, neutral, or I don't know), with document titles presented in random order (Figure \ref{fig:evaluation}). For Comprehension, they judged whether four factual and four counterfactual statements from the documents were right or wrong (right, wrong, or I don't know), presented in random order. We sequenced these components so that earlier questions wouldn't inadvertently provide clues to later answers.}

\edit{Finally, participants rated their agreement with statements about critical thinking components, adapted from the Critical Thinking Self-Assessment Scale \cite{payan2022development, tanprasert2024debate}, followed by feedback on the study.}

\subsection{Dependent Variables}\label{DVs}

In the critical thinking performance assessment task, participants typically go through the process of browsing documents for an overview, selecting key documents to examine closely, drawing and inferring evidence-based arguments, and synthesizing these arguments into an essay for their reasoned decision \cite{braun2020performance,shavelson2019assessment}. While the framework provides an authentic measure of critical thinking performance \cite{braun2020performance,shavelson2019assessment}, participants may not explicitly present all information they process in their final essays due to individual judgment about what to include. Therefore, we supplement the core Essay performance with additional measures to capture related cognitive activities: recall of information provided, evaluation of sources, and comprehension through basic factual inference. These four components of performance---Essay, Recall, Evaluation, and Comprehension---provide a comprehensive assessment of critical thinking task performance and related cognitive activities.

\subsubsection{Essay}\label{essay_scoring}
Participants' essays, as the product of their reasoning, were graded using the scoring scheme \cite{EbrightJonesCortina2025,ebright2024we} of the performance assessment task \cite{braun2020performance,shavelson2019assessment}. \edit{This scoring scheme emphasizes simplicity through its arithmetic count-based approach, applied in previous research demonstrating improvement in critical thinking performance before and after training \cite{EbrightJonesCortina2025}.} 
It showed strong content, criterion, and construct validity: it measures a representative sample of critical thinking subskills, scores correspond to established indicators, and it captures the full scope of critical thinking processes \cite{EbrightJonesCortina2025}. 
Based on an exhaustive list of pre-identified valid arguments that can be inferred from the constructed set of documents \cite{EbrightJonesCortina2025,ebright2024we}, we awarded points for: valid arguments drawn from the documents ($+$1 per argument matching the pre-identified list), explicit references to source documents ($+$1 per reference), and explicit evaluation of source trustworthiness ($+$1 per evaluation). Penalties were applied for fabrications ($-$1), writing errors ($-$1), and repeated arguments ($-$1). The total Essay score sums these components, capturing both breadth (quantity of valid arguments) and depth (argumentation quality). Additionally, for assessing reasoning balance, the Myside Bias score captures whether participants incorporate multiple perspectives or exhibit one-sided reasoning \cite{ebright2024we,stanovich2008failure,wolfe2008locus,stanovich2007natural}, calculated as the absolute difference between pro and con arguments. 

\begin{itemize}
    \item \textbf{Essay score}: Sum of valid arguments ($+$1 each), document references ($+$1 each), trustworthiness evaluations ($+$1 each), minus fabrications ($-$1 each), writing errors ($-$1 each), and repeated arguments ($-$1 each).
    \item \textbf{Number of arguments}: Sum of valid arguments ($+$1 each).
    \item \textbf{Myside Bias score}: Absolute difference between number of valid pro and con arguments.
\end{itemize}

\subsubsection{Recall}
Recall assesses participants' internalization and retention of the source documents. Participants' free recall answers summarizing the main ideas of documents were scored against the main idea for each document (Table \ref{tab:document_overview}). Each correctly summarized document received 1 point, partially correct summaries received 0.5 points, and incorrect or missing summaries received 0 points.
\begin{itemize}
\item \textbf{Recall score}: Sum across documents (1 = correct, 0.5 = partial, 0 = incorrect/missing).
\end{itemize}

\subsubsection{Evaluation}
Evaluation assesses participants' judgments of source characteristics independent of what they included in their essays.
For each document, participants either rated its relevance, trustworthiness, and stance, or selected ``I didn't read this.'' Ratings were scored against the actual characteristics of each document (Table \ref{tab:document_overview}). Correct ratings received 1 point, while incorrect ratings, ``I don't know'' or ``I didn't read this'' received 0 points.

\begin{itemize}
\item \textbf{Evaluation (overall)}: Average percentage of documents correctly evaluated across relevance, trustworthiness, and stances.
\item \textbf{Evaluation (relevance)}: Percentage of documents correctly evaluated for relevance.
\item \textbf{Evaluation (trustworthiness)}: Percentage of documents correctly evaluated for trustworthiness.
\item \textbf{Evaluation (stances)}: Percentage of documents correctly evaluated for stance.
\end{itemize}

\subsubsection{Comprehension}
Comprehension assesses participants' understanding of factual evidence from the documents.
Participants' answers judging four factual and four counterfactual statements were scored against the actual correctness of each statement (see Supplementary Materials for the correct answers). Correct judgments received 1 point, while incorrect judgments or ``I don't know'' received 0 points.

\begin{itemize}
\item \textbf{Comprehension}: Percentage of correctly judged statements.
\end{itemize}

\subsection{Study Interface}

We created a web interface for experiment participants to complete the main performance assessment task and answer all the questions. Below, we describe how the interface works for the main task (Section \ref{task_interface}) and the LLM-powered chatbot used for providing LLM access (Section \ref{llm_chatbot}).

\subsubsection{Main task interface}\label{task_interface}

Figure \ref{fig:interface} shows the interface used for the main task in the six conditions with LLM access (\earlyai, \continuousai and \lateai LLM access under \insufficient and \sufficient time availability). The interface consists of three panels: a document viewer for reading source documents (left), the LLM-powered chatbot (right), and an essay entry textbox (bottom). The interface dynamically controls LLM access by enabling and disabling the chatbot based on the assigned time constraints. Timed alerts appear at designated points (when three minutes and one minute remain) to notify participants about upcoming changes in LLM access and remaining time. A countdown timer displays remaining time, and participants are automatically advanced to the next page when time expires. Participants with \noai LLM access had only the document viewer and essay entry textbox.

\subsubsection{LLM-powered chatbot}\label{llm_chatbot}

We employed a widely-used general purpose model at the time (OpenAI's GPT-4o) without substantial modification. Given the constructed set of documents in the critical thinking performance assessment task, we created an LLM-powered chatbot that operates exclusively on these documents without internet connectivity.
The chatbot had all documents pre-loaded, including titles, sources, and text descriptions of images and tables. When participants entered queries, the chatbot responded based on the queries and the documents, with references linked to the document viewer.


\edit{We designed the system prompt directing the model to answer queries based on the documents, refining it iteratively through multiple rounds of pilots before the main experiment, refining the model behavior to be representative of real-world human-LLM interaction. This process led us to key insights, for example, we found that adding ``refer to the files you think that are most useful to answer the question'' improved response quality by enabling query-relevant retrieval rather than uniform reference to all documents. In our testing and pilots, we found no instance of the model misrepresenting content from source documents or citing incorrect documents, though errors may occur with atypical queries.}

\subsection{Participants}

\edit{For the preregistered experiment,} we recruited 400 participants from Prolific. We had 393 participants in all after excluding those who refreshed the web interface during the study. Our study interface allowed one submission for each. Based on a power analysis using pilot and simulated data, this sample size is needed for 80\% power, a medium effect size, using a significance level of 0.05. \edit{The 393 participants came from diverse races, genders, ages, and education groups (Table \ref{tab:demographics} in Appendix).} For \insufficient time availability, the overall study took around 20 minutes to complete, and each participant was paid \$4. For \sufficient time availability, the overall study took around 40 minutes to complete, and each participant was paid \$8. Payment was based on an hourly rate of \$12. 
Participants had a 100\% approval rate on Prolific, were based in the US, and fluent in English. The study was approved by the Institutional Review Board (IRB) of our institution. 

Our experiment yielded 51 with \noai LLM access, 50 with \continuousai LLM access, 39 with \earlyai LLM access, and 50 with \lateai LLM access under \insufficient time availability; 52 with \noai LLM access, 52 with \continuousai LLM access, 51 with \earlyai LLM access, and 48 with \lateai LLM access under \sufficient time availability. \edit{Sample sizes were balanced across LLM access timing conditions ($\chi^2(3) = 1.07$, $p = 0.79$) and time availability conditions ($\chi^2(1) = 0.43$, $p = 0.51$).}

\subsection{Analysis}

\subsubsection{Grading process}

\edit{The scoring scheme for Essay performance (Section \ref{essay_scoring}) makes the primary grading task a binary identification of whether each valid argument is present. Given that LLMs can provide reliable and consistent annotation \cite{xiao2023supporting, rathje2024gpt, matter2024close} and grading of writing quality \cite{rathje2024gpt, hackl2023gpt, lira2025learning} that closely match human judgments, we followed the practice of using an LLM to grade essays at scale, providing it with the scoring scheme, the list of valid arguments with index numbers, all task documents, and instructions to output the indices of identified arguments and count each score component. See Supplementary Materials for prompt and model specifications.}

\edit{To confirm reliability and consistency, one author manually scored a stratified sample of 100 participants across conditions. Agreement between human and LLM grading was assessed at two levels. For participants' essays, the correlation coefficient ICC(2,1) values ranged from 0.70 to 0.82 for count-based score components, all above the recommended threshold of 0.70 \cite{gisev2013interrater,hallgren2012computing}; for binary identification of each valid argument's presence, percentage agreement averaged 94.7\% with mean Cohen's $\kappa$ of 0.76. We also graded all the LLM responses from the experiment to assess argument overlap, validated using the LLM responses in this stratified sample. Reliability was similarly strong: ICC(2,1) values ranged from 0.78 to 0.84, with 95.5\% agreement and mean Cohen's $\kappa$ of 0.81 for binary identification. See Supplementary Materials for detailed statistics.}

\subsubsection{Statistical analysis}
Following our pre-registration,\footnote{https://aspredicted.org/q6yh-mrqg.pdf} we conducted an Analysis of Covariance (ANCOVA) analysis to examine the main effects and interaction of LLM access timing and time availability on all dependent variables. The model included time availability (\insufficient, \sufficient) and LLM access timing (\earlyai, \continuousai, \lateai, \noai LLM access) as fixed factors, adjusting for participants' LLM use frequency, LLM attitudes, self-efficacy in critical thinking, and confidence in LLM capabilities. All tests used a significance level of 0.05. We conducted Tukey's Honestly Significant Difference (HSD) post-hoc tests for pairwise comparisons. Comparisons were performed within each time availability level, comparing all pairs of LLM access timing conditions, and within each LLM access timing condition, comparing time availability.

\section{Results}\label{main_results}

In this section, we present the results from our pre-registered analyses. We begin with a summary of our key findings using the main measures for Essay, Recall, Evaluation, and Comprehension (Section \ref{results_summary}). We then present detailed results of all measures for Essay (Section \ref{essay_results}), Recall (Section \ref{recall_results}), Evaluation (Section \ref{evaluation_results}), and Comprehension (Section \ref{comprehension_results}).

\begin{figure*}[htbp]
    \centering
    \begin{minipage}[t]{0.5\textwidth}
        \centering
        \begin{subfigure}[t]{0.95\textwidth}
            \centering
            \includegraphics[width=\linewidth]{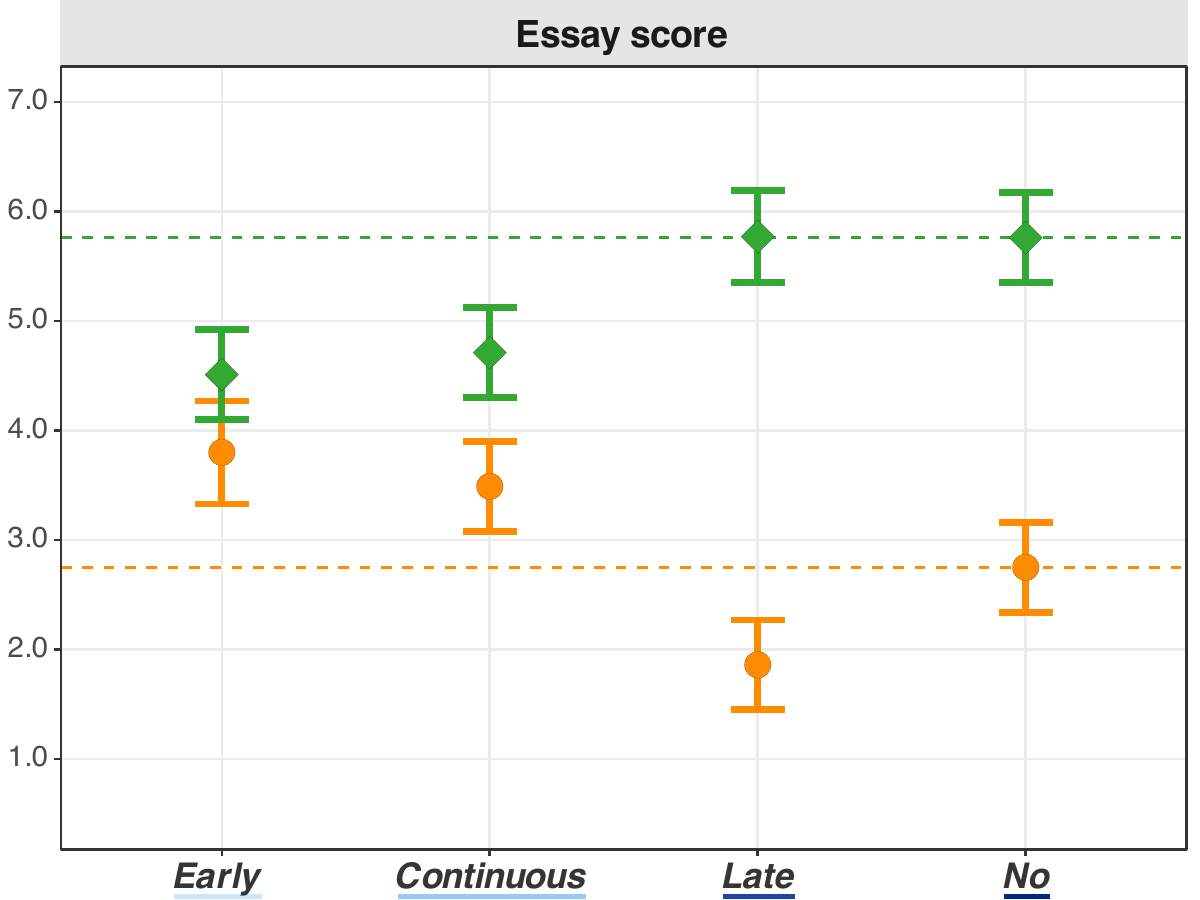}
            \caption{Essay score: patterns reversed based on time availability. Under \insufficient time, participants having LLM access from the start (\earlyai, \continuousai LLM access) outperformed those who worked independently first (\lateai, \noai LLM access). Notably, those having LLM access from the start showed minimal gains from \insufficient to \sufficient time. See Section \ref{essay_score}.}
            \label{fig:essay-score}
        \end{subfigure}
    \end{minipage}%
    \begin{minipage}[t]{0.5\textwidth}
        \centering
        \begin{subfigure}[t]{0.95\textwidth}
            \centering
            \includegraphics[width=\linewidth]{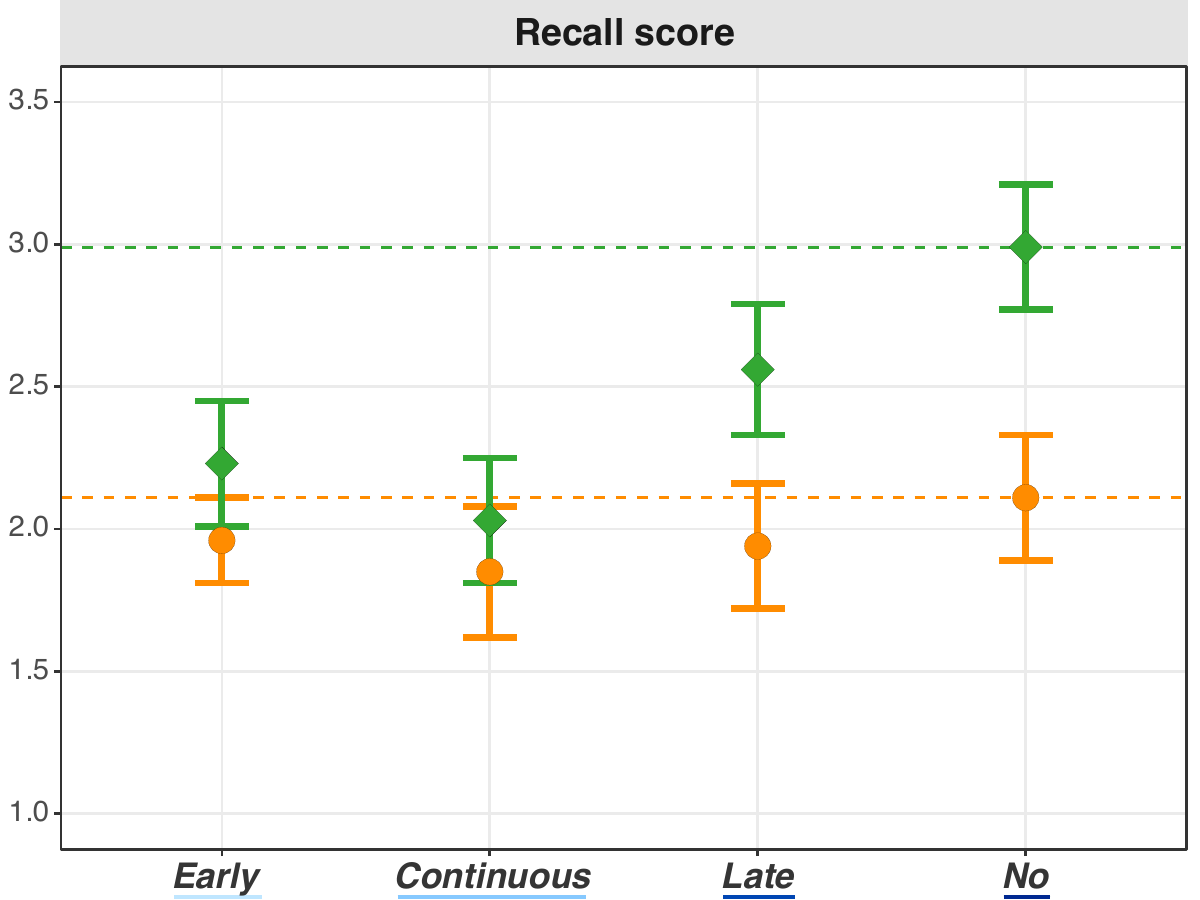}
            \caption{Recall score: under \insufficient time, all LLM access timings showed similarly poor Recall. Under \sufficient time, having LLM access from the start (\earlyai, \continuousai LLM access) impaired Recall compared to working independently first (\lateai, \noai LLM access). See Section \ref{recall_results}.}
            \label{fig:literal-recall}
        \end{subfigure}
    \end{minipage}
    
    \vspace{0.5em}
    
    \begin{minipage}[t]{0.5\textwidth}
        \centering
        \begin{subfigure}[t]{0.95\textwidth}
            \centering
            \includegraphics[width=\linewidth]{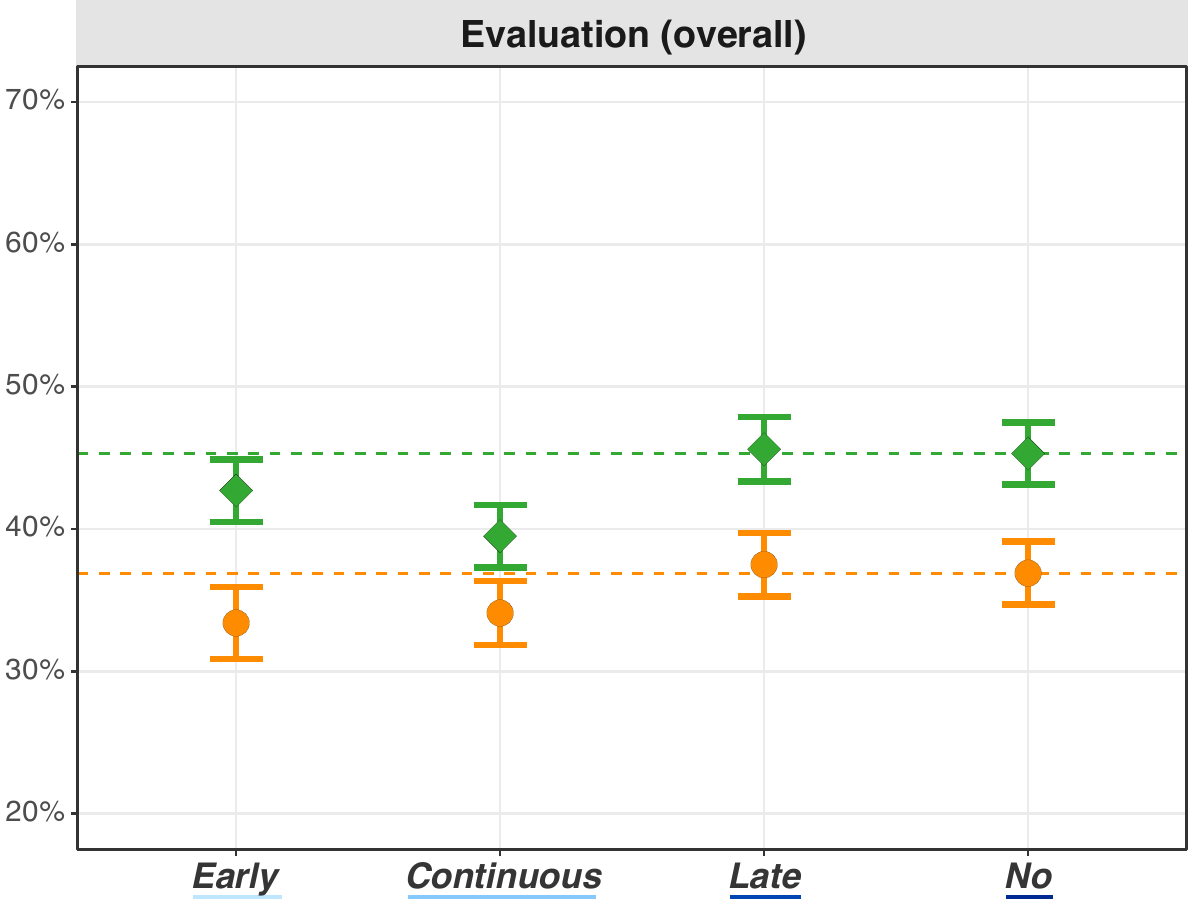}
            \caption{Evaluation (overall): \sufficient time generally improved Evaluation correctness overall, with minimal differences between LLM access timings. See Section \ref{evaluation_results}.}
            \label{fig:overall-correctness}
        \end{subfigure}
    \end{minipage}%
    \begin{minipage}[t]{0.5\textwidth}
        \centering
        \begin{subfigure}[t]{0.95\textwidth}
            \centering
            \includegraphics[width=\linewidth]{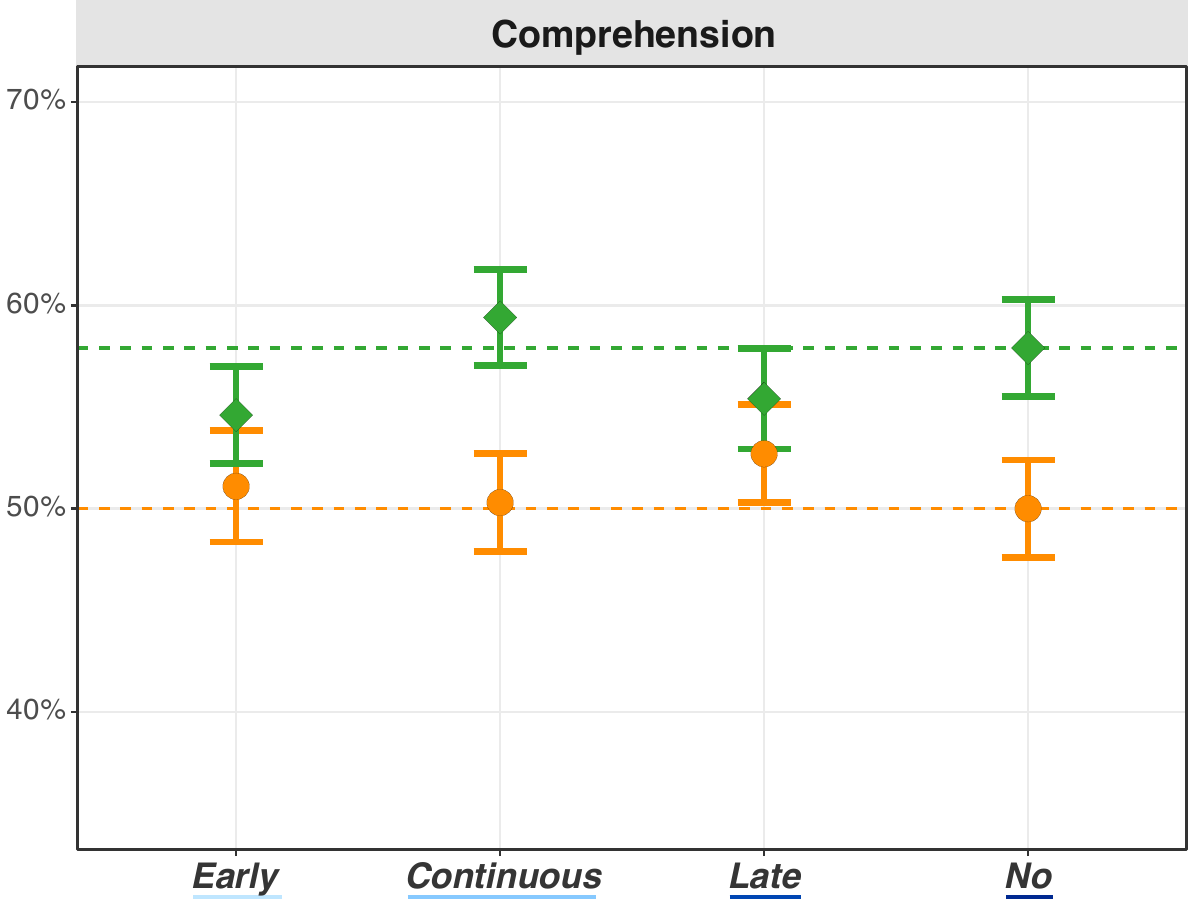}
            \caption{Comprehension: minimal sensitivity to LLM access timing, and modest improvement from \insufficient to \sufficient time. See Section \ref{comprehension_results}.}
            \label{fig:comprehension}
        \end{subfigure}
    \end{minipage}
    
    \caption{Estimated means of the main score for each measure across all conditions. Circles and diamonds show the estimated means when time availability is \insufficient and \sufficient, respectively. Error bars represent standard errors. Dashed lines indicate the reference of \noai LLM access. The effects of LLM access timing and time availability were most pronounced for (a) Essay score and (b) Recall score. See Table \ref{tab:ancova-posthoc-insufficient} and \ref{tab:ancova-posthoc-sufficient} for detailed pairwise comparisons between LLM access timings.}
    \label{fig:main-results}
\end{figure*}


\subsection{Summary of Results}\label{results_summary}

As shown in Figure \ref{fig:main-results}, the effects of LLMs on critical thinking task performance were not uniform but dependent on temporal context. 
For \textbf{RQ1} regarding the effect of LLM access timing under different time availability: 

\begin{itemize}
  \mainfinding{1}{Under \insufficient time, having LLM access from the start (\earlyai and \continuousai LLM access) improved participants’ \textnormal{Essay} performance of critical thinking, compared to working independently first (\lateai and \noai LLM access), suggesting that LLMs can provide an initial boost under time pressure (Section~\ref{essay_score}).}
  \mainfinding{2}{Under \sufficient time, participants having LLM access from the start (\earlyai and \continuousai LLM access) showed a trend toward lower \textnormal{Essay} performance, compared to those working independently first (\lateai and \noai LLM access), suggesting that late LLM access can preserve critical thinking when time allows deliberation (Section~\ref{essay_score}).}
  \mainfinding{3}{Under \sufficient time, having \lateai LLM access reduced Myside Bias, compared to having \noai LLM access, while maintaining a similar number of arguments (Section~\ref{myside_bias_score}).}
  \mainfinding{4}{Under \sufficient time, having LLM access from the start (\earlyai and \continuousai LLM access) reduced participants’ \textnormal{Recall} of information provided, compared to having \noai LLM access, suggesting that having LLM access from the start can prevent participants from internalizing information even when time permits deep engagement (Section~\ref{recall_results}).}
\end{itemize}

For \textbf{RQ2} regarding the effect of time availability when having the same LLM access timing:

\begin{itemize}
  \mainfinding{5}{Having \sufficient time largely improved participants’ \textnormal{Essay} performance of critical thinking for those working independently first (\lateai and \noai LLM access), and only modestly improved for those having LLM access from the start (\earlyai and \continuousai LLM access; Section~\ref{essay_score}).}
  \mainfinding{6}{Having \sufficient time largely improved participants’ \textnormal{Recall} performance for those working independently first (\lateai and \noai LLM access), but not for those having LLM access from the start (\earlyai and \continuousai LLM access; Section~\ref{recall_results}). }
  \mainfinding{7}{Having \sufficient time generally improved participants’ \textnormal{Evaluation} (Section~\ref{evaluation_results}) and \textnormal{Comprehension} (Section~\ref{comprehension_results}).}
\end{itemize}

\begin{table*}[htbp]
\centering
\small
\renewcommand{\arraystretch}{1}
\setlength{\tabcolsep}{2pt}
\begin{tabular*}{\textwidth}{@{\extracolsep{\fill}}l l c c c c l@{}}
\toprule
\textbf{Sec.} & \textbf{Measures} & \textbf{\earlyai LLM access} & \textbf{\continuousai LLM access} & \textbf{\lateai LLM access} & \textbf{\noai LLM access} & \textbf{Post-hoc analysis} \\
\midrule
\multirow[c]{3}{*}{\textbf{\ref{essay_results}}}
& Essay score & 3.80 (0.47) & 3.49 (0.41) & 1.86 (0.41) & 2.75 (0.41) &
\begin{tabular}[t]{@{}l@{}}\earlyai $>$ \lateai LLM access (*)\\ \continuousai $>$ \lateai LLM access (*)\end{tabular} \\
\cmidrule(lr{0pt}){2-7}
& Number of arguments & 3.72 (0.30) & 3.26 (0.27) & 2.58 (0.27) & 3.03 (0.27) &
\earlyai $>$ \lateai LLM access (*) \\
\cmidrule(lr{0pt}){2-7}
& Myside Bias score & 3.13 (0.30) & 2.63 (0.27) & 2.03 (0.27) & 2.49 (0.26) &
\earlyai $>$ \lateai LLM access (*) \\
\midrule
\multirow[c]{1}{*}{\textbf{\ref{recall_results}}}
& Recall score & 1.96 (0.15) & 1.85 (0.23) & 1.94 (0.22) & 2.11 (0.22) & \\
\midrule
\multirow[c]{4}{*}{\textbf{\ref{evaluation_results}}}
& Evaluation (overall) & 33.4\% (2.52) & 34.1\% (2.24) & 37.5\% (2.23) & 36.9\% (2.21) & \\
\cmidrule(lr{0pt}){2-7}
& Evaluation (relevance) & 39.4\% (3.28) & 38.6\% (2.91) & 44.2\% (2.89) & 42.6\% (2.87) & \\
\cmidrule(lr{0pt}){2-7}
& Evaluation (trustworthiness) & 28.5\% (3.03) & 28.3\% (2.69) & 35.6\% (2.68) & 35.3\% (2.65) & \\
\cmidrule(lr{0pt}){2-7}
& Evaluation (stances) & 32.2\% (3.15) & 35.6\% (2.79) & 32.8\% (2.78) & 32.8\% (2.75) & \\
\midrule
\multirow[c]{1}{*}{\textbf{\ref{comprehension_results}}}
& Comprehension & 51.1\% (2.73) & 50.3\% (2.42) & 52.7\% (2.41) & 50.0\% (2.39) & \\
\bottomrule
\end{tabular*}
\caption{Summaries of results comparing LLM access timings under \insufficient time availability. We report the estimated means (standard errors). The last column shows significant post-hoc differences (+ $p{<}0.1$, * $p{<}0.05$, ** $p{<}0.01$, *** $p{<}0.001$).}
\label{tab:ancova-posthoc-insufficient}
\end{table*}


\begin{table*}[htbp]
\centering
\small
\renewcommand{\arraystretch}{1}
\setlength{\tabcolsep}{2pt}
\begin{tabular*}{\textwidth}{@{\extracolsep{\fill}}l l c c c c l@{}}
\toprule
\textbf{Sec.} & \textbf{Measures} & \textbf{\earlyai LLM access} & \textbf{\continuousai LLM access} & \textbf{\lateai LLM access} & \textbf{\noai LLM access} & \textbf{Post-hoc analysis} \\
\midrule
\multirow[c]{3}{*}{\textbf{\ref{essay_results}}}
& Essay score & 4.51 (0.41) & 4.71 (0.41) & 5.77 (0.42) & 5.76 (0.41) & \\
\cmidrule(lr{0pt}){2-7}
& Number of arguments & 4.56 (0.26) & 4.42 (0.26) & 5.25 (0.27) & 5.12 (0.26) & \\
\cmidrule(lr{0pt}){2-7}
& Myside Bias score & 3.79 (0.26) & 3.34 (0.26) & 3.42 (0.27) & 4.41 (0.26) &
\begin{tabular}[t]{@{}l@{}}\continuousai $<$ \noai LLM access (*)\\ \lateai $<$ \noai LLM access (*)\end{tabular} \\
\midrule
\multirow[c]{1}{*}{\textbf{\ref{recall_results}}}
& Recall score & 2.23 (0.22) & 2.03 (0.22) & 2.56 (0.23) & 2.99 (0.22) &
\begin{tabular}[t]{@{}l@{}}\earlyai $<$ \noai LLM access (+)\\ \continuousai $<$ \noai LLM access (*)\end{tabular} \\
\midrule
\multirow[c]{4}{*}{\textbf{\ref{evaluation_results}}}
& Evaluation (overall) & 42.7\% (2.20) & 39.5\% (2.19) & 45.6\% (2.28) & 45.3\% (2.19) & \\
\cmidrule(lr{0pt}){2-7}
& Evaluation (relevance) & 53.6\% (2.86) & 45.3\% (2.84) & 52.7\% (2.96) & 50.8\% (2.85) & \\
\cmidrule(lr{0pt}){2-7}
& Evaluation (trustworthiness) & 37.7\% (2.65) & 34.0\% (2.63) & 39.0\% (2.74) & 44.3\% (2.63) &
\continuousai $<$ \noai LLM access (*) \\
\cmidrule(lr{0pt}){2-7}
& Evaluation (stance) & 36.7\% (2.75) & 39.3\% (2.73) & 45.1\% (2.84) & 44.4\% (2.73) & \\
\midrule
\multirow[c]{1}{*}{\textbf{\ref{comprehension_results}}}
& Comprehension & 54.6\% (2.38) & 59.4\% (2.37) & 55.4\% (2.47) & 57.9\% (2.37) & \\
\bottomrule
\end{tabular*}
\caption{Summaries of results comparing LLM access timings under \sufficient time availability. We report the estimated means (standard errors). The last column shows significant post-hoc differences (+ $p{<}0.1$, * $p{<}0.05$, ** $p{<}0.01$, *** $p{<}0.001$).}
\label{tab:ancova-posthoc-sufficient}
\end{table*}

\begin{figure*}[htbp]
  \centering
  \begin{minipage}[t]{0.5\textwidth}
    \centering
    \begin{subfigure}[t]{0.95\textwidth}
      \centering
      \includegraphics[width=\linewidth]{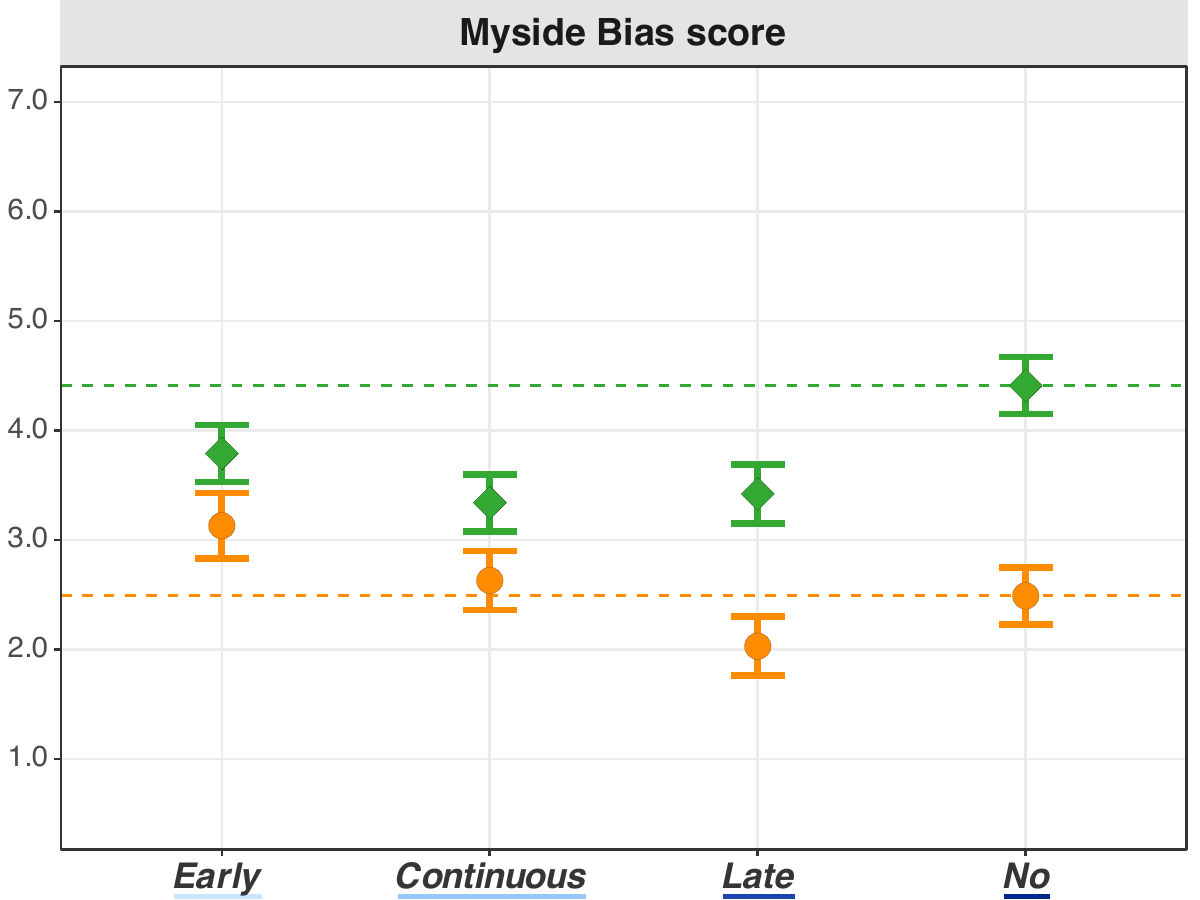}
      \caption{Myside Bias score: under \sufficient time, \lateai LLM access reduced Myside Bias compared to \noai LLM access.}
      \label{fig:bias}
    \end{subfigure}
  \end{minipage}%
  \begin{minipage}[t]{0.5\textwidth}
    \centering
    \begin{subfigure}[t]{0.95\textwidth}
      \centering
      \includegraphics[width=\linewidth]{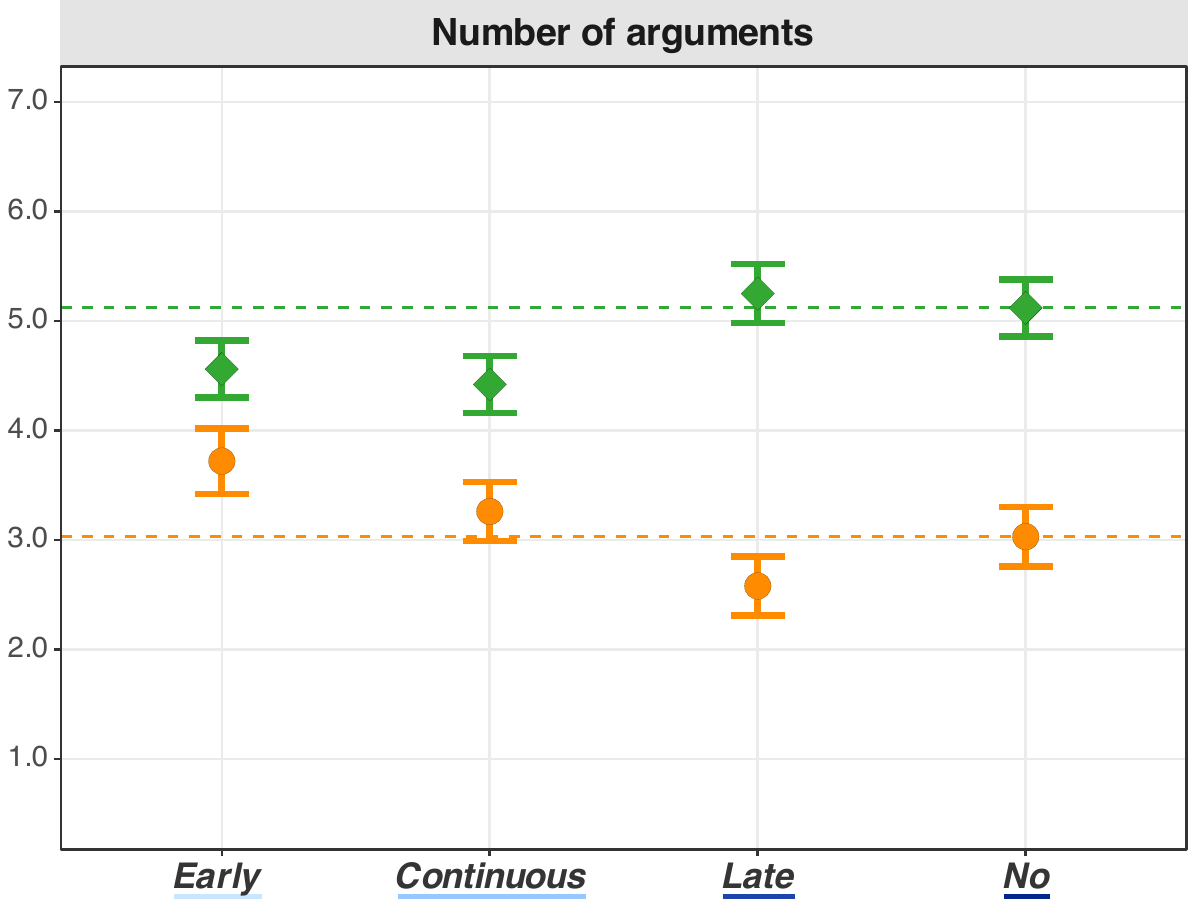}
      \caption{Number of arguments: under \sufficient time, \lateai and \noai LLM access showed similar argument quantity.}
      \label{fig:args}
    \end{subfigure}
  \end{minipage}
  \caption{Estimated means of (a) Myside Bias score and (b) number of arguments across conditions. Error bars represent standard errors. Dashed lines indicate the reference of \noai LLM access. Comparing (a) and (b), only \lateai LLM access appeared to reduce Myside Bias while maintaining argument quantity under \sufficient time, whereas the stable low Myside Bias for \earlyai and \continuousai LLM access may reflect fewer arguments rather than balanced reasoning. See Section \ref{myside_bias_score}.}
  \label{fig:bias-args}
\end{figure*}




\subsection{Essay}\label{essay_results}

\subsubsection{Essay score}\label{essay_score}
For the Essay score, the ANCOVA analysis revealed a significant interaction between LLM access timing and time availability ($F(3, 381) = 6.39$, $p < 0.001$) and a main effect of time availability ($F(1, 381) = 54.78$, $p < 0.001$), indicating that the effects of LLM access timing depended on time availability. We detailed the patterns from the post-hoc analysis as follows. 


\paragraph{Effects of LLM access timing by time availability.} The effects of LLM access timing are substantially dependent on whether participants had \insufficient or \sufficient time availability (Figure \ref{fig:essay-score}). 

\textbf{Under insufficient time, having LLM access from the start improved Essay performance.} With \insufficient time availability, participants having LLM access from the start outperformed those who attempted the task independently first, suggesting that using LLMs to start off the critical thinking task can provide initial scaffolding that boosts the Essay performance. Table \ref{tab:ancova-posthoc-insufficient} shows the trend that, when time availability was \insufficient, \earlyai access yielded the highest Essay score ($M = 3.80$), followed by \continuousai LLM access ($M = 3.49$), \noai LLM access ($M = 2.75$), and \lateai LLM access showing the poorest ($M = 1.86$). Post-hoc analysis showed significant advantages for having \earlyai over \lateai LLM access ($p < 0.01$), and having \continuousai over \lateai LLM access ($p < 0.05$). 

\textbf{Under sufficient time, working independently first showed better Essay performance.} With \sufficient time availability, overall, having LLM access for the critical thinking task has little benefit compared to not having LLM access. Participants with LLM access from the start showed a trend toward lower Essay scores, and those working independently first showed a similar Essay score to those without LLM access. Table \ref{tab:ancova-posthoc-sufficient} shows the trend that, with \sufficient time availability, \lateai ($M = 5.77$) and \noai LLM access ($M = 5.76$) exhibited higher Essay score than \continuousai ($M = 4.71$) and \earlyai LLM access ($M = 4.51$). 


\paragraph{Effects of time availability by LLM access timing.} While having more time availability generally improved Essay performance, this benefit varied dramatically across LLM access timings (Figure \ref{fig:essay-score}).

\textbf{Sufficient time was substantially beneficial for participants working independently first.} For those having \lateai and \noai LLM access, the benefit of \sufficient time over \insufficient time was substantially larger (\lateai: 3.91 points, $p < 0.001$ and \noai: 3.01 points, $p < 0.001$), suggesting that working independently first allows participants to benefit from more time for deliberation.

\textbf{Sufficient time had little benefit for participants having LLM access from the start.} For participants having \earlyai LLM access, those under \sufficient time showed minimal improvement in Essay score compared to those under \insufficient time. For participants having \continuousai LLM access, the benefit of \sufficient time over \insufficient time was significant but modest (1.22 points, $p < 0.05$). This suggests that giving LLMs access from the start may prematurely constrain participants’ thinking, preventing them from the benefits of more time for deliberation.

The number of valid arguments, as the primary component of the Essay score, showed a similar trend (see Table \ref{tab:ancova-posthoc-insufficient} and \ref{tab:ancova-posthoc-sufficient}). These results suggest that, under \insufficient time having LLM access from the start can provide crucial scaffolding for rapid argument generation, but with \sufficient time, the same scaffolding appears to limit participants’ thoughts for optimal critical thinking task performance that they could achieve independently. Under \insufficient time, \lateai LLM access may come too late to help, but with \sufficient time, it can best preserve and augment participants’ thoughts after initial independent work.

\subsubsection{Myside Bias score}\label{myside_bias_score}

The ANCOVA analysis revealed significant main effects of time availability ($F(1, 381) = 36.52$, $p < 0.001$) and LLM access timing ($F(3, 381) = 3.58$, $p < 0.05$), with a marginal interaction ($F(3, 381) = 2.55$, $p < 0.1$). 

\textbf{Late LLM access under sufficient time reduced Myside Bias while maintaining argument quantity.} Under \sufficient time (Table \ref{tab:ancova-posthoc-sufficient} and Figure \ref{fig:bias-args}), participants having \lateai and \noai LLM access produced similar numbers of arguments ($M=5.25$ and $M=5.12$), yet the former ($M=3.42$) showed significantly lower Myside Bias than the latter ($M=4.41$; $p<0.05$). Under \insufficient time (Table \ref{tab:ancova-posthoc-insufficient} and Figure \ref{fig:bias-args}), participants having \lateai LLM access ($M=2.03$) exhibited lower Myside Bias than those having \earlyai LLM access Myside Bias ($M=3.13$; $p<0.05$), but this mirrors argument quantity: participants having \lateai LLM access ($M=2.58$) produced fewer arguments than those having \earlyai LLM access ($M=3.72$; $p<0.05$). 

For the effects of time availability, participants with \noai LLM access showed a 1.92-point increase in Myside Bias from \insufficient to \sufficient time, suggesting that long-term independent thinking can entrench one-sided reasoning. While participants having \lateai LLM access showed a 1.39-point increase from \insufficient to \sufficient time, they exhibited significantly less Myside Bias than those having \noai LLM access ($p<0.05$). This suggests that \lateai LLM access can effectively reduce the one-sided reasoning that develops during independent work.


In contrast, under \sufficient time, participants having \earlyai ($M=3.79$) and \continuousai LLM access ($M=3.34$) showed lower Myside Bias than those having \noai LLM access, but produced the fewest arguments (Table \ref{tab:ancova-posthoc-sufficient} and Figure \ref{fig:bias-args}). Thus, their stable Myside Bias scores reflected limited argumentation rather than genuine consideration of diverse perspectives---echoing our earlier finding that having LLM access from the start can anchor users to initial thoughts.

\begin{figure*}[htbp]
  \centering
  \begin{subfigure}[t]{0.31\textwidth}
    \centering
    \includegraphics[width=\linewidth]{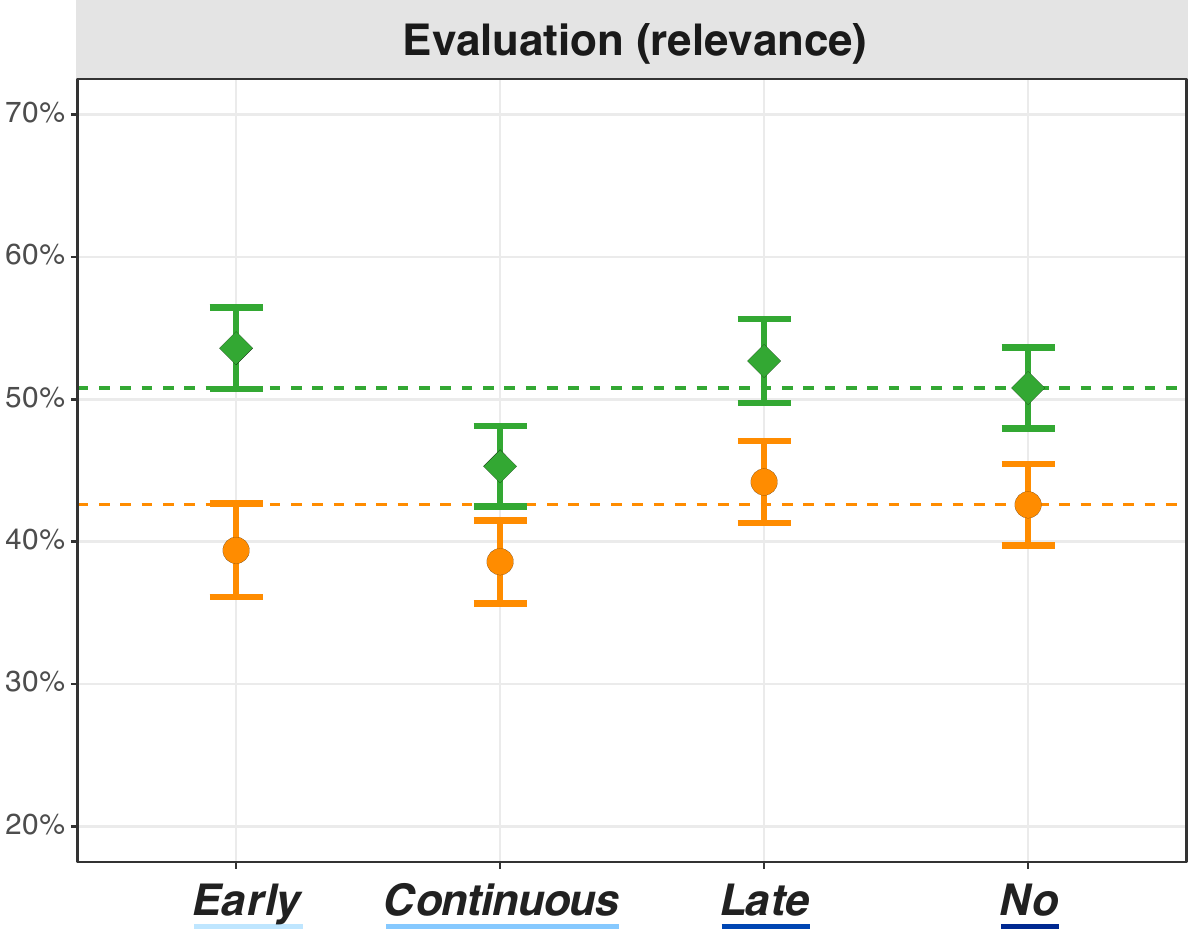}
    \caption{Evaluation (relevance)}
    \label{fig:relevance}
  \end{subfigure}
  \hfill
  \begin{subfigure}[t]{0.31\textwidth}
    \centering
    \includegraphics[width=\linewidth]{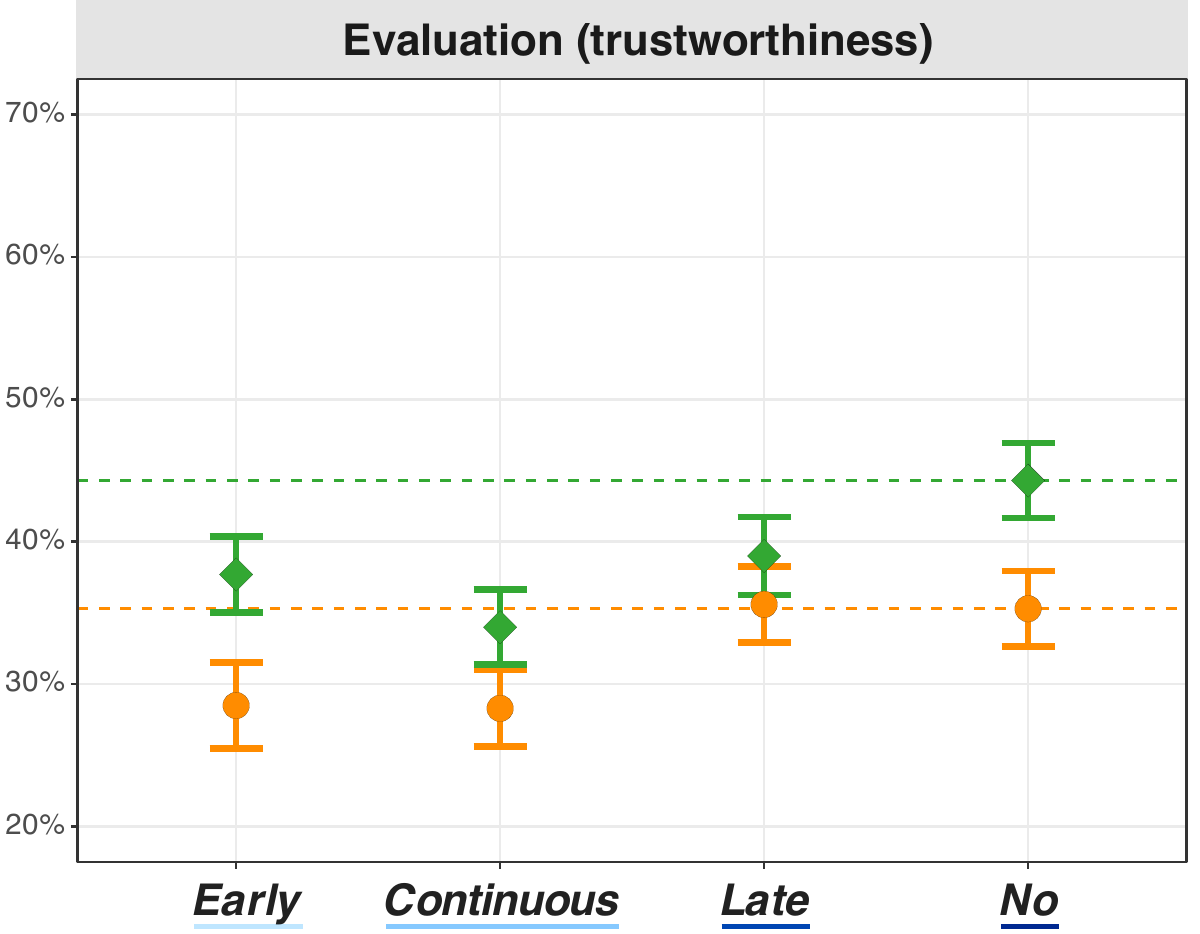}
    \caption{Evaluation (trustworthiness)}
    \label{fig:trustworthiness}
  \end{subfigure}
  \hfill
  \begin{subfigure}[t]{0.31\textwidth}
    \centering
    \includegraphics[width=\linewidth]{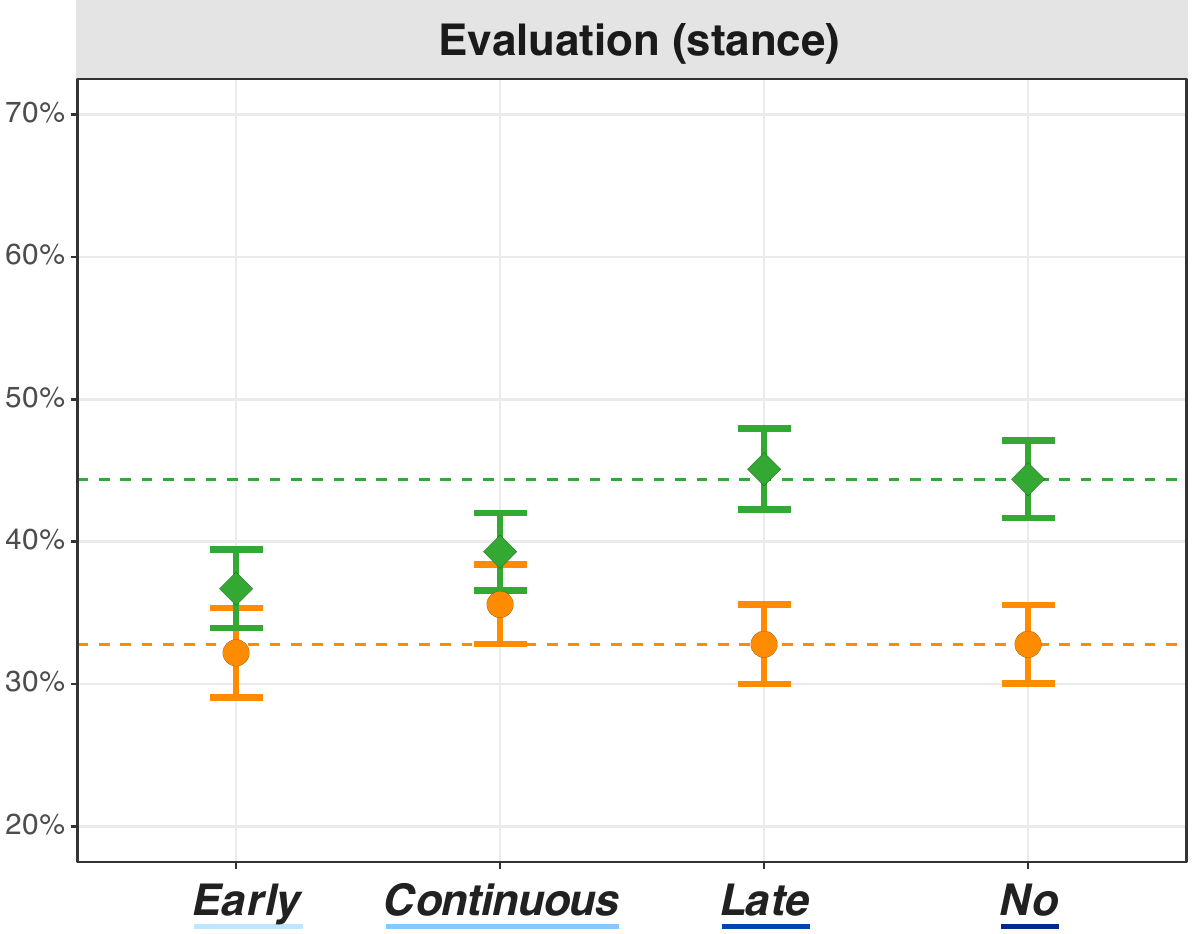}
    \caption{Evaluation (stances)}
    \label{fig:stance}
  \end{subfigure}
  \caption{Estimated means of Evaluation for all three components across conditions. Error bars represent standard errors. Dashed lines indicate the reference of \noai LLM access. \textit{Sufficient} time generally improved Evaluation with modest differences between LLM access timings. The exception was Evaluation (trustworthiness), where \continuousai LLM access impaired it under \sufficient time. See Section \ref{evaluation_results}.}
  \label{fig:evaluation-1x3}
\end{figure*}

\subsection{Recall}\label{recall_results}

The ANCOVA analysis revealed significant main effects of LLM access timing ($F(3, 380) = 2.79$, $p < 0.05$) and time availability ($F(1, 380) = 9.07$, $p < 0.01$), indicating that both factors influenced participants’ Recall performance during the task.

\paragraph{Effects of LLM access timing by time availability.}
Table \ref{tab:ancova-posthoc-insufficient} shows that, under \insufficient time, there were no significant differences in Recall score between LLM access timings.
\textbf{Under sufficient time, having LLM access from the start impaired Recall performance.} Table \ref{tab:ancova-posthoc-sufficient} shows that, in Recall score, participants having \earlyai ($M = 2.23$) and \continuousai LLM access ($M = 2.03$) scored lower than those having \noai LLM access ($M = 2.99$; $p < 0.1$ and $p < 0.05$ respectively). Meanwhile, participants having \lateai LLM access ($M = 2.56$) scored between these extremes.
This pattern suggests that participants having LLM access from the start may not internalize information, even when time permits deep engagement with the documents.

\paragraph{Effects of time availability by LLM access timing.} \textbf{Sufficient time substantially improved Recall performance of participants working independently first.} Participants having \noai LLM access showed significant improvement in Recall score from \insufficient to \sufficient time (a 42\% increase; $p < 0.01$). Similarly, those having \lateai LLM access improved substantially from \insufficient to \sufficient time (a 32\% increase; $p < 0.01$). In contrast, participants having \earlyai and \continuousai LLM access showed minimal improvement. This reinforces that having LLM access from the start may not only limit participants' thoughts but also memory retention with the documents.

\subsection{Evaluation}\label{evaluation_results}

The ANCOVA analysis showed that time availability significantly affected overall Evaluation performance ($F(1, 381) = 23.29$, $p < 0.001$) and each component (relevance: $F(1, 381) = 19.62$, $p < 0.001$; trustworthiness: $F(1, 381) = 12.40$, $p < 0.001$; stance: $F(1, 381) = 13.00$, $p < 0.001$). \textit{Sufficient} time generally improved overall Evaluation performance (Figure \ref{fig:overall-correctness}) and each component (Figure \ref{fig:evaluation-1x3}). On the other hand, LLM access timing only affected the trustworthiness component ($F(3, 381) = 4.34$, $p < 0.01$). The modest effect size and small pairwise differences observed should be interpreted with caution. 
In pairwise comparisons, under \sufficient time, for Evaluation (trustworthiness), participants having \continuousai LLM access ($M = 34.0\%$) performed significantly worse than those having \noai LLM access ($M = 44.3\%$; $p < 0.05$), with \earlyai ($M = 37.7\%$) and \lateai LLM access ($M = 39.0\%$) falling in between. This exception may indicate that continuous LLM access reduced attention to source credibility, as participants having \continuousai LLM access may rely on the LLM heavily rather than developing their own calibration.

\subsection{Comprehension}\label{comprehension_results}

The ANCOVA analysis revealed a significant main effect of time availability ($F(1, 381) = 11.11$, $p<0.001$), with no significant effect of LLM access timing or interaction. 
The benefit of having more task time was significant for participants having \continuousai ($p<0.01$) and \noai LLM access ($p<0.05$; Figure \ref{fig:comprehension}). Overall, Comprehension showed minimal sensitivity to LLM access timing.

\subsection{Self-Reported Assessment of Critical Thinking}\label{self_assessment}

The self-assessment showed slight variation across conditions (Figure \ref{fig:self_assessment_plots} in Appendix), with only time availability showing some difference. This contrasts sharply with the substantial performance differences observed, highlighting the necessity of performance assessment to capture the nuances of LLMs' impact on critical thinking, especially under time constraints. 

\section{Behavioral Engagement}\label{behavior_engagement}


To better understand the mechanisms underlying the main analysis results, we examined how participants behaviorally engaged with the LLM access and documents based on their interaction logs and self-reports. We report observations on textual overlap (Section \ref{textual_overlap}); argument overlap (Section \ref{argument_overlap}) between participants and the LLM responses they received; how participants having \lateai LLM access might revise their essays (Section \ref{late_ai_reduce_mysidebias}); document viewing behavior (Section \ref{doc_viewing}); and participants’ task approaches (Section \ref{task_approach}). Main observations are summarized as follows: 



\begin{itemize}
  \item{\textbf{Limited unique arguments when having LLM access from the start:} participants having LLM access from the start (\earlyai and \continuousai LLM access) showed minimal increase in unique arguments (non-overlapping with LLM responses) from \insufficient to \sufficient time, potentially explaining their minimal Essay performance improvement (Section \ref{argument_overlap}). }
  \item{\textbf{Late LLM access for balancing perspectives:} participants having \lateai LLM access could incorporate arguments on the other side during the \lateai LLM access period, potentially explaining their reduced Myside Bias (Section \ref{late_ai_reduce_mysidebias}).}
  \item{\textbf{Reduced document iteration when having LLM access from the start:} participants having LLM access from the start (\earlyai and \continuousai LLM access) showed minimal increase in documents viewed during writing from \insufficient to \sufficient time, while those working independently first (\lateai and \noai LLM access) largely increased (Section \ref{doc_viewing}).}
  \item{\textbf{Distinct task approaches:} under \insufficient time, participants having \earlyai LLM access primarily used the LLM access for document summarization (35.9\%) while those having \lateai LLM access often couldn't effectively use the LLM access (38.0\%); under \sufficient time, participants having \lateai LLM access employed the LLM access for confidence and reassurance (31.3\%; Section \ref{task_approach}).}
\end{itemize}


\subsection{Textual Overlap with LLM Responses}\label{textual_overlap}

\begin{table*}[htbp]
\centering
\small
\renewcommand{\arraystretch}{1}
\begin{tabular*}{\textwidth}{@{\extracolsep{\fill}}l c c c c c c c@{}}
\toprule
& & \multicolumn{3}{c}{\textbf{Copying}} & \multicolumn{3}{c}{\textbf{Textual overlap}} \\
\cmidrule(lr){3-5} \cmidrule(lr){6-8}
\textbf{Condition} & \textbf{N} & \textbf{Instances (\%)} & \textbf{Copied wordcount} & \textbf{Rate (\%)} & \textbf{Instances (\%)} & \textbf{Overlapped wordcount} & \textbf{Rate (\%)} \\
\midrule
\insufficient time & & & & & & & \\
\quad \earlyai LLM access & 39 & 3 (8\%) & 39.7 (13.7) & 20.6 (15.6) & 28 (72\%) & 75.9 (69.1) & 38.8 (20.6) \\
\quad \continuousai LLM access & 50 & 3 (6\%) & 119.7 (41.5) & 57.5 (14.2) & 32 (64\%) & 92.9 (80.9) & 50.7 (29.6) \\
\quad \lateai LLM access & 50 & 2 (4\%) & 140.0 (42.4) & 49.3 (11.8) & 35 (70\%) & 66.7 (87.0) & 39.5 (25.3) \\
\quad \noai LLM access & 51 & 0 (0\%) & -- & -- & 0 (0\%) & -- & -- \\
\midrule
\sufficient time & & & & & & & \\
\quad \earlyai LLM access & 51 & 7 (14\%) & 37.3 (42.9) & 10.0 (7.3) & 48 (94\%) & 110.0 (70.2) & 36.0 (20.9) \\
\quad \continuousai LLM access & 52 & 6 (12\%) & 63.2 (47.0) & 19.8 (12.3) & 44 (85\%) & 147.5 (103.5) & 49.2 (24.7) \\
\quad \lateai LLM access & 48 & 7 (15\%) & 39.9 (34.6) & 16.1 (16.1) & 36 (75\%) & 89.8 (41.6) & 34.2 (13.2) \\
\quad \noai LLM access & 52 & 0 (0\%) & -- & -- & 0 (0\%) & -- & -- \\
\bottomrule
\end{tabular*}
\caption{Copying and textual overlap between participants' essays and LLM responses across conditions. Instances show count (percentage) of participants. Rate is the copied/overlapped wordcount divided by the essay wordcount, shown as mean (standard deviation) percentage. Copied/overlapped wordcount and rate are among those with copying/overlap instances.}
\label{tab:textual_overlap}
\end{table*}

\begin{figure*}[htbp]
  \centering
  \includegraphics[width=0.95\textwidth]{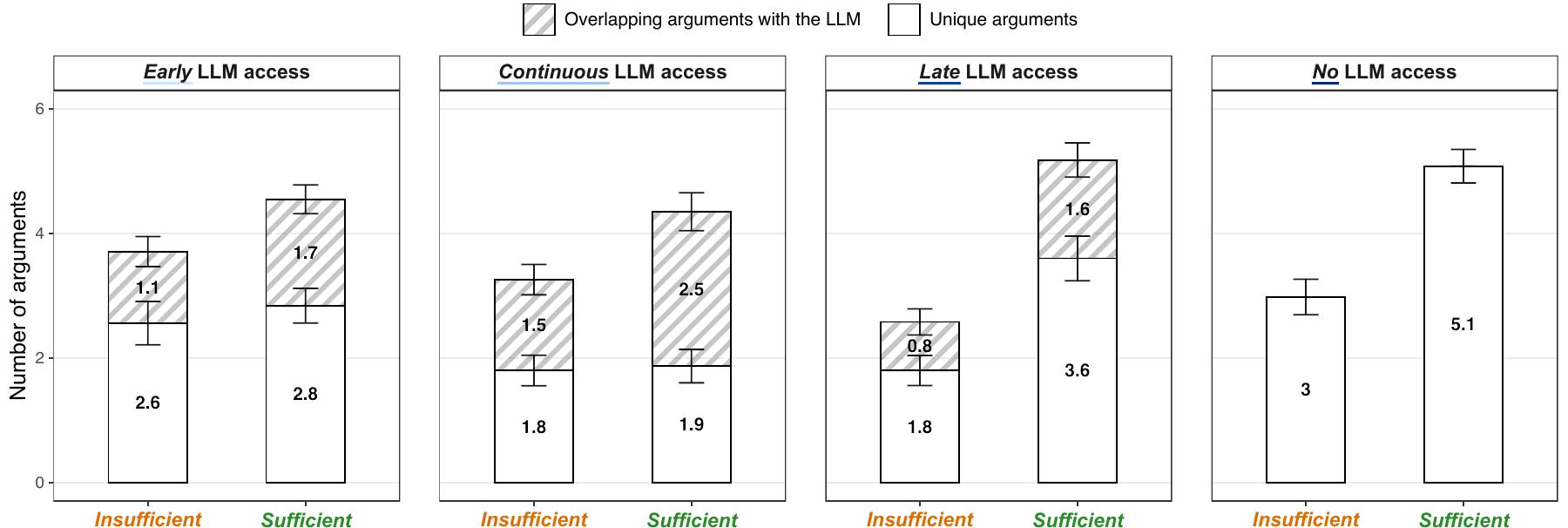}
  \caption{Overlap between participant arguments and LLM responses across conditions. Striped bars indicate overlapping arguments; solid bars indicate unique (non-overlapping) arguments. Error bars represent standard errors. From \insufficient to \sufficient time, participants working independently first (\lateai and \noai LLM access) developed notable more unique arguments, while those having LLM access from the start (\earlyai and \continuousai LLM access) did not.}
  \label{fig:argument_overlap}
\end{figure*}


We checked participants' copying behavior and textual overlap with the LLM to understand how participants incorporated LLM responses into their essays. 
Textual overlap measures lexical similarity. For each participants, we split both their essay and the LLM responses they received into sentences, then broke each sentence into words. For each sentence in the participant's essay, we identified the LLM sentence with the largest word overlap and counted the number of shared words. We summed these counts across all sentences in the participant's essay to get the overlapped wordcount, and divided by the total wordcount to obtain a rate. For explicit copying, we logged copy-paste events, verified that copied fragments originated from the LLM responses, identified matching words in the participant's essay, and divided by the total wordcount.

Table \ref{tab:textual_overlap} shows that direct copying was rare across all conditions, but textual overlap was common (64--94\% of participants with LLM access), suggesting that participants might not directly copy from but have been influenced by LLM responses. Among textual overlap instances, participants under \sufficient time showed higher average overlapped wordcount than those under \insufficient time. The average overlapped wordcount was highest for \continuousai LLM access (\insufficient: 92.9; \sufficient: 147.5) and lowest for \lateai LLM access (\insufficient: 66.7; \sufficient: 89.8). It is not surprising for \continuousai LLM access, as these participants had LLM access throughout the task, indicating higher reliance. Conversely, the lower overlap for \lateai LLM access suggests that participants working independently first could develop their essays more in their own words, reducing subsequent direct incorporation of LLM responses.

\subsection{How Might Participants Develop Arguments?}

First, beyond textual overlap, we checked the overlap between valid arguments in participants' essays and LLM responses (Section \ref{argument_overlap}). This serves as an indicator of how much argumentative content participants may have drawn from the LLM, though participants may have independently arrived at similar arguments. Second, we explored how participants having \lateai LLM access under \sufficient time might have used the LLM to reduce Myside Bias (Section \ref{late_ai_reduce_mysidebias}).



\subsubsection{Argument overlap with LLM responses}\label{argument_overlap}

Participants having LLM access from the start (\earlyai and \continuousai LLM access) were exposed to more valid arguments on average in the LLM responses than those having \lateai LLM access, under both \insufficient (\earlyai: 2.72; \continuousai: 3.52; \lateai: 2.28) and \sufficient time (\earlyai: 4.16; \continuousai: 6.73; \lateai: 3.65). However, this greater exposure did not translate to more arguments in participants' essays under \sufficient time. A notable pattern emerged (Figure~\ref{fig:argument_overlap}): for participants having LLM access from the start, their unique arguments showed negligible increase from \insufficient to \sufficient time (\earlyai: 0.28; \continuousai: 0.07). In contrast, participants working independently first developed more unique arguments from \insufficient to \sufficient time (\lateai: 1.80; \noai: 2.10). This indicates that, while participants having LLM access from the start (\earlyai and \continuousai LLM access) were exposed to more valid arguments in the LLM responses and they had more overlapped arguments with the LLM, they appeared limited in their further deliberation.


\begin{figure*}[htbp]
    \centering
    \includegraphics[width=1\textwidth]{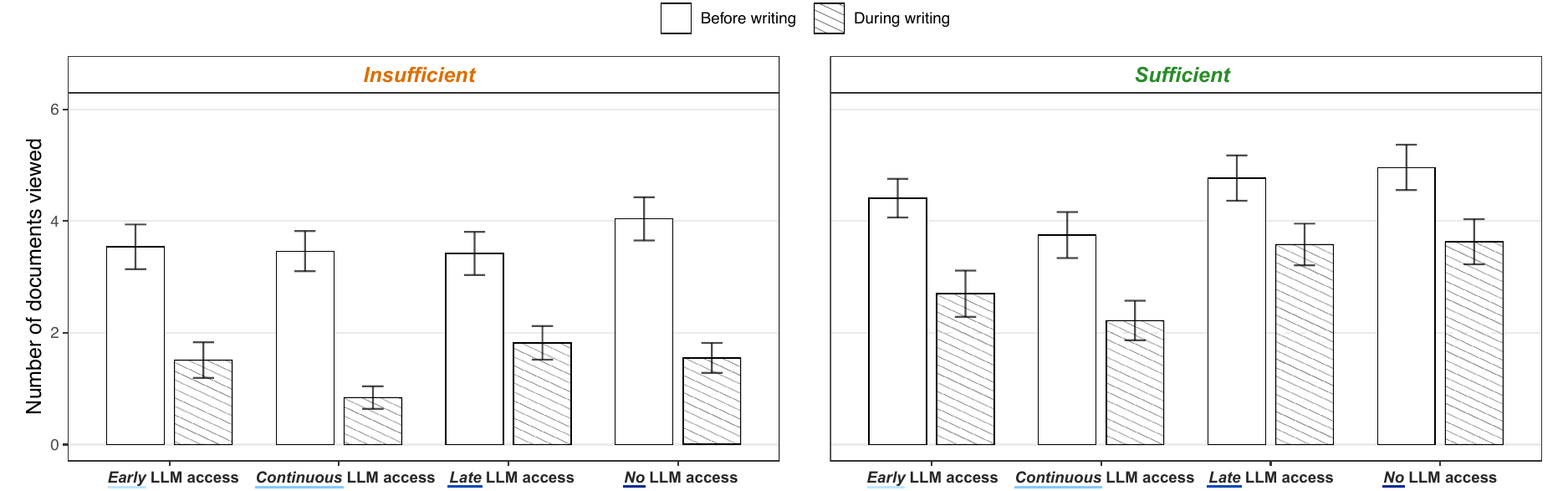}
    \caption{Number of unique documents participants viewed before and during writing across conditions. Error bars represent standard errors. 
    From \insufficient to \sufficient time, document viewing increased during writing for \lateai and \noai LLM access, indicating iteration. This trend was modest for \earlyai and \continuousai LLM access.}
    \label{fig:doc_viewing}
\end{figure*}

This may help explain the reversed trend of Essay performance (Section \ref{essay_results}): having LLM access from the start (\earlyai and \continuousai LLM access)may have facilitated rapid argumentation but limited further deliberation even when more time was available. In particular, participants with \earlyai LLM access did not appear to benefit from the independent working time after LLM access was removed, despite having the same independent time as those with \lateai LLM access.





\definecolor{highlight}{rgb}{1,1,0.85}  
\sethlcolor{highlight}

\subsubsection{How might late LLM access help reduce Myside Bias?}\label{late_ai_reduce_mysidebias}

An interesting finding from our main results is that participants with \lateai LLM access showed significantly reduced Myside Bias compared to those with \noai LLM access under \sufficient time, while maintaining similar argument quantity (Section \ref{myside_bias_score}).
Thus, we examined how participants having \lateai LLM access revised their essays during the access period by comparing two versions of each participant's essay: the one recorded just before the access period began and the final submitted version. 

Under \insufficient time, participants made modest revisions by adding only 0.68 arguments on average during the late LLM access period, suggesting that the LLM access might arrive too late to help. Under \sufficient time, however, \lateai LLM access appeared valuable for reducing Myside Bias. Of the 48 participants in this condition, 15 reduced their Myside Bias during the late LLM access period. Notably, 8 who initially presented only con arguments added an average of 2.22 pro arguments that also appeared in the LLM responses. For example, P31 (having \lateai LLM access under \sufficient time) explained \edit{in post-task open-ended response explaining their task approaches}: \textit{``I read through the documents myself. I then made my decision and wrote the essay. I asked the AI Chatbot if I should accept Hallman's offer, the pros and cons, to strengthen my argument from what the AI gave me.''} Their query to the LLM was straightforward: \textit{``Should the city accept or reject Hallman's \$19 million offer?''} The LLM response present several pros and cons, \edit{including the following highlighted pro argument reasoning about the financial benefits of accepting Hallman's offer:} 
\begin{quote}
\textit{``[...] Here are the pros and cons based on the documents: Pros: \hl{Reduces the financial burden on the city for future liabilities outside the contamination above 50 ppb since Hallman would take over these responsibilities. Offers a clear end to ongoing obligations, possibly saving costs and focusing resources on remediation.}''}
\end{quote}
\edit{This was not in P31's essay before the LLM access period, but was in the final essay. The highlighted text shows what P31 added:} 
\begin{quote}\textit{``[...] Since exafluoran is a toxic chemical that is highly flammable, has a faint pleasant smell and is toxic for rare microbes we would not want to risk lives. \hl{The pros of accepting the offer is it will offer an immediate financial relief, resolve the issue without delay, offer a clear end to ongoing litigations and saving costs}. [...]''}
\end{quote}

This suggests participants could use late LLM access as a final check and quality assurance to consider overlooked perspectives, which potentially explain why \lateai LLM access reduced Myside Bias compared to \noai LLM access under \sufficient time.





\begin{table*}[htbp]
\centering
\small
\renewcommand{\arraystretch}{1}
\setlength{\tabcolsep}{4pt}
\begin{tabular*}{\textwidth}{@{}l|l|l|l@{}}
\toprule
\textbf{Condition} & \textbf{Most common approach (\%)} & \textbf{Second most common approach (\%)} & \textbf{Third most common approach (\%)} \\
\midrule
\textbf{\insufficient time} & & & \\
\quad \earlyai LLM access & AI for Summarizing Documents (35.9) & AI for Content Adoption (20.5) & AI for Verification \& Clarification (20.5) \\
\quad \continuousai LLM access & AI for Verification \& Clarification (28.0) & AI for Content Adoption (26.0) & AI for Summarizing Documents (22.0) \\
\quad \lateai LLM access & Minimal/No AI Use (38.0) & AI for Verification \& Clarification (22.0) & AI for Confidence \& Reassurance (10.0) \\
\quad \noai LLM access & Systematic Document Review (51.0) & Selective Document Review (27.5) & Time Management Strategy (11.8) \\
\midrule
\textbf{\sufficient time} & & & \\
\quad \earlyai LLM access & AI for Verification \& Clarification (35.3) & AI for Summarizing Documents (27.5) & AI for Content Adoption (23.5) \\
\quad \continuousai LLM access & AI for Verification \& Clarification (37.8) & AI for Summarizing Documents (20.3) & AI for Structuring Pros \& Cons (13.5) \\
\quad \lateai LLM access & AI for Confidence \& Reassurance (31.3) & AI for Verification \& Clarification (27.1) & AI for Structuring Pros \& Cons (20.8) \\
\quad \noai LLM access & Systematic Document Review (50.0) & Heuristic Judgments (18.8) & Time Management Strategy (16.2) \\
\bottomrule
\end{tabular*}
\caption{Most common task approaches across all conditions (top 3 most frequent approaches). See Appendix \ref{description_task_approach} for full descriptions.}
\label{tab:top_task_approaches}
\end{table*}

\subsection{Document Viewing Behavior}\label{doc_viewing}

As the critical thinking task involves a non-linear process of reading documents and writing essays, we examined participants' document viewing behavior before and during writing to understand how participants might deliberate in forming and refining their thoughts.
As expected, participants under \sufficient time generally viewed more unique documents than those under \insufficient time, particularly during writing (Figure \ref{fig:doc_viewing}).
This suggests that when time permitted, participants engaged with the documents iteratively during writing, potentially returning to sources to verify and strengthen their arguments.

Notably, LLM access timing revealed distinct document-viewing patterns. Participants having LLM access from the start (\continuousai and \earlyai LLM access) viewed fewer documents than those who worked independently first (\lateai and \noai LLM access). This trend was most pronounced for document viewing during writing under \sufficient time: for example, participants having \continuousai LLM access checked fewer documents (2.42 on average) during writing compared to participants having \noai LLM access (3.63 on average).


These patterns reinforce our earlier findings that participants having LLM access from the start (\continuousai and \earlyai LLM access) showed poorer Essay and Recall performance under \sufficient time. Their reduced document engagement---particularly the lack of iterative consultation during writing---suggests they might become anchored to their initial thoughts, rather than further iterate with the source documents.


\subsection{Participants' Task Approaches}\label{task_approach}

To qualitatively understand participants' task approaches in their conditions, we analyzed their post-task open-ended responses explaining their approaches. We focused on their open-ended responses rather than queries, as queries may not reflect actual intentions. For example, P31's query did not reveal their use of the LLM to consider diverse perspectives (Section \ref{late_ai_reduce_mysidebias}). Table \ref{tab:top_task_approaches} shows the most common approaches across conditions.

Under \insufficient time, ``AI for Summarizing Documents'' (using the LLM access to summarize documents) was most common for participants having \earlyai LLM access (35.9\%), while ``AI for Content Adoption'' (directly incorporating LLM-generated text into essays) was most common for participants having \continuousai LLM access (26.0\%). Notably, ``Minimal/No AI Use'' emerged as the most common approach for participants having \lateai LLM access (38.0\%), indicating they might not find the \lateai LLM access helpful under time pressure. 

Under \sufficient time, ``AI for Verification \& Clarification" (using the LLM access to confirm facts and resolve uncertainties) was the primary approach for both participants having \earlyai (35.3\%) and \continuousai LLM access (37.8\%). For participants having \lateai LLM access, ``AI for Confidence \& Reassurance'' (using the LLM access to validate already-formed arguments) was the most common approach (31.3\%),  followed by ``AI for Verification \& Clarification" (27.1\%), and ``AI for Structuring Pros \& Cons'' (using the LLM access to organize and balance perspectives; 20.8\%), suggesting strategically employing the LLM to strengthen what they already had.

\section{Discussion}

\subsection{Key Findings}

\subsubsection{LLM access after independent work: preservation and augmentation when time permits}

The effects of \lateai LLM access were largely dependent on the time availability. Under \insufficient time, \lateai LLM access provided minimal benefit for the Essay score, as the critical thinking task performance. Specifically, participants may not make effective use of the LLM. 
However, when time permits more deliberation, \lateai LLM access best preserved Essay performance compared to \noai LLM access, while \earlyai and \continuousai LLM access showed a trend toward worse performance. Participants with \lateai LLM access iterated with more documents during writing and could use the LLM as a post-hoc check to strengthen their thinking. This pattern echoes the idea that working independently first allowed participants to develop their schema \cite{bo2025s,tankelevitch2025understanding}, which, in turn, enabled more strategic LLM use.

Our findings shed light on scenarios of making reasoned decisions about new problems, in which participants formed judgments without pre-existing opinions or background knowledge. This offers a different perspective on concerns about LLMs creating ``echo chambers'' that reinforce pre-existing opinions \cite{sharma2024generative,hart2009feeling}. Independent work as a solo deep dive may naturally lead to more entrenched one-sided reasoning over time; Myside bias is a natural tendency when given no instructions to avoid it \cite{stanovich2008failure,wolfe2008locus,stanovich2007natural}. However, \lateai LLM access could counteract this tendency: it reduced Myside bias compared to \noai LLM access under \sufficient time while maintaining argument quantity, as participants could incorporate overlooked viewpoints as a final check.

\subsubsection{LLM access from the start: boost under time pressure, impairment under sufficient time}

Having LLM access from the start created a trade-off between immediate efficiency and deep deliberation. Under time pressure, participants with \earlyai and \continuousai LLM access showed the best Essay performance. When time was \sufficient, this pattern reversed. 
Behaviorally, they developed fewer unique arguments and iterated less with documents during writing compared to those who worked independently first, suggesting that they might be constrained by their initial thoughts. This echoes the idea that AI outputs can act as anchors, increasing confidence in initial AI-recommended directions while reducing motivation for continued exploration and independent assessment \cite{rastogi2022deciding,nguyen2024human}.

Notably, most participants who had LLM access from the start did not directly offload decision-making or the task---they used the LLM access to summarize or clarify document content. Yet this seemingly intentional use of LLMs for initial navigation appeared sufficient to anchor users' subsequent deliberation about which documents, ideas, or stances to pursue. This raises important considerations for reflective AI use and metacognitive awareness \cite{tankelevitch2024metacognitive,lee2025impact}. Even not naive use of LLMs---for initial navigation rather than direct offloading of task completion---can limit downstream deliberation of a complex reasoning task. Beyond verifying the accuracy of LLM outputs, users also need to assess the diversity and completeness of the sources LLMs present.

\smallskip

Taken together, our findings reveal a dramatic temporal reversal of LLM access timing effects under different time availability: what helps under time pressure---having LLM access from the start---can harm when time permits deliberation, and the benefits of having more time for deliberation---typically taken for granted---do not hold. This challenges the simple narratives about LLMs being universally beneficial or harmful for reasoning. Optimal human-AI collaboration for critical thinking tasks requires careful consideration of when and under what time constraints LLMs should be introduced, not simply whether they should be available.

\subsection{\edit{Design Implications for Using LLMs for Critical Thinking} }

\edit{Recent work studying how users enact critical thinking when using GenAI provided implications to help users think critically about GenAI use \cite{lee2025impact}. Our work, studying whether and when LLMs support or undermine critical thinking performance, provides insights for a fundamentally different question: how to support users when they use LLMs for tasks that require critical thinking. Our findings highlight the need for temporally-aware LLM applications. The implications can also apply to information-seeking and online reasoning scenarios that require critical thinking.}

\subsubsection{\edit{Encouraging independent work before LLM use}}

\edit{Our findings suggest that with sufficient time, users having late LLM access showed optimal performance, leveraging the LLM to strengthen what they had developed. For better critical thinking outcomes, LLM tools could encourage users to attempt tasks independently first, for example with nudging messages like \emph{``what are your initial thoughts before I help?''}~\cite{caraban201923}.
Interventions could also employ mechanisms to delay LLM assistance for tasks requiring critical thinking. One approach is friction-based interventions \cite{silva2022strategies,mejtoft2019design,collins2024modulating}, such as slowing response speed to give users time to formulate their own thoughts before receiving LLM assistance.}

\edit{However, applying such interventions might frustrate users seeking quick assistance with routine tasks. Therefore, it is important to identify tasks that would benefit. Recent work analyzing user conversations with LLMs classified tasks by the required cognitive skills \cite{handa2025economic}. Such classification could help detect tasks that require critical thinking and selectively apply interventions.}

\edit{Users with strong self-regulation or domain experts might be able to proactively delay LLM access and use it more intentionally \cite{dziak2025srt,bo2025s}, yet many users may not.
For unfamiliar tasks, early LLM guidance can be necessary,
but this guidance should focus on orientation and navigation rather than solution generation, preserving opportunities for independent reasoning. Some recent designs provide promising starting points. ChatGPT's study mode \cite{openai_study_mode}, for example, prompts users with guiding questions rather than providing direct answers. While users have to manually opt into such modes now, we envision that LLM applications could automatically detect critical thinking tasks and proactively suggest the use of these modes to users.}


\subsubsection{\edit{Supporting LLM use from the start under time pressure}}
\edit{In practice, users routinely face deadlines from external demands or internal expectations \cite{hoge2009work,baethge2018matter,zuzanek2004work,kelly1988entrainment}. Under such time pressure, users may turn to LLMs for productivity boosts \cite{noy2023experimental,bick2025impact,spatharioti2025effects}. Our findings suggest that providing LLM access from the start can indeed offer advantages over requiring users to work independently first. However, early LLM access also risks anchoring users' subsequent deliberation, shaping which documents they attend to, which ideas they pursue, and which stances they consider. Design interventions should therefore mitigate this risk when supporting rapid assistance under time pressure. Furthermore, rather than solely prompting users to verify output quality, interventions should target deeper aspects of critical thinking, such as source diversity and reasoning balance. For example, metacognitive prompts encouraging users to engage more thoroughly with sources and consider alternative perspectives have shown promise in LLM-assisted search tools \cite{singh2025protecting}. Also, LLM tools might present multiple candidate outputs grounded in different subsets of sources or visualize metrics of source diversity and credibility, enabling time-pressured users to assess informational balance without substantial additional effort.}


\subsubsection{Adapting LLM access timing to real-world time availability}

\edit{Ideally, users would allocate abundant time for reasoning and make use of LLMs after engaging in independent work, which our findings suggest can support stronger critical thinking outcomes.} However, in reality, users do not always manage their time optimally---procrastination, unexpected deadlines, and urgent demands are common in professional and educational settings \cite{steel2022self,prem2018procrastination}. 

Design should account for these imperfect temporal realities. Rather than offering static support, LLM applications should adapt dynamically to users' time availability, which raises key design questions: What counts as sufficient versus insufficient time for a given task? When should LLM access be introduced? And how should assistance change when users are rushed versus when they have time to think? For complex reasoning tasks, designers must consider not only whether to provide LLM support, but when and how to tailor assistance as time progresses. 
\edit{For instance, LLM applications can incorporate lightweight time-management features that ask users how much time they have when they begin a task and adjust support accordingly. Such tools can encourage independent work when time is plentiful, or offer immediate scaffolding when users are under pressure. Embedded timers could track session duration to enable time-aware guidance, and browser extensions could extend such capabilities across multiple LLM applications in a workflow.}

\edit{Furthermore, it is important to raise public understanding about how LLM use affects cognitive processes under different temporal conditions, consistent with broader goals in AI literacy \cite{kasinidou2022promoting,schuller2022data}. Recent meta-analyses have examined when human-AI collaboration is beneficial across various tasks \cite{vaccaro2024combinations}, yet time constraints---which our work highlights as pivotal---are largely absent from these accounts. Expanding such research and translating it into accessible resources, such as searchable databases or interactive tutorials \cite{zhao2025thinking}, could help users develop better metacognition about when and how to use LLMs given their own temporal realities, and potentially foster better time management.}

\subsubsection{\edit{Designing for both immediate performance and preserving cognitive capacities}}

When designing LLM applications for critical thinking, it is crucial to consider their long-term impact on human cognitive and learning capacities beyond immediate performance. 
Our results show that having LLM access from the start impaired participants' memory retention of the provided documents, even when time was sufficient, suggesting that early LLM access may disrupt natural cognitive pathways and the internalization of knowledge that support critical thinking.

Researchers have argued that without regular practice of applying internalized information for higher-order thinking, individuals may lose these capacities over time \cite{oakley2025memory}. Also, cognitive offloading by relying on external tools like ChatGPT may disrupt the brain's natural learning systems \cite{kosmyna2025your}. Thus, designers and educators should inform users to periodically practice cognitive activities without LLM assistance to help maintain these skills. 

Designing for both immediate efficiency and skill maintenance requires deliberate choices about when and how LLM assistance becomes available. 
Organizations should navigate the trade-off between immediate task performance and longer-term capability development. In educational settings, a university-provided LLM application could adapt its functionality based on remaining time before a deadline---limiting early access to navigation support, then unlocking features for higher-order thinking as students progress and deadlines approach. In professional contexts, such as developers navigating existing codebases to make architectural decisions, LLM applications could unlock only after users document their preliminary approach, preserving independent practice while positioning LLMs to identify edge cases.


\subsection{Implications for Human-AI Collaboration Research} 

\subsubsection{Performance assessment}

Our findings highlight the importance of performance assessment for measuring thinking and learning outcomes. Most notably, self-report measures showed minimal variation, while performance assessment revealed substantial differences in critical thinking task performance under different time constraints. We argue that self-reports may miss crucial nuances in how LLMs affect cognitive processes. Given that LLMs' impact on critical thinking is shaped by time constraints, participants may have blind spots about subtle cognitive changes in different temporal contexts.
Other research also showed that people have poor awareness of their own time use \cite{collopy1996biases,parry2021systematic}, and self-reports lack the resolution to capture real-time dynamics of cognitive processes \cite{pekrun2020self}. For evaluation of human-AI collaboration on thinking and learning, authentic performance tasks should be prioritized.


\begin{figure*}[htbp]
  \centering
  \includegraphics[width=0.95\textwidth]{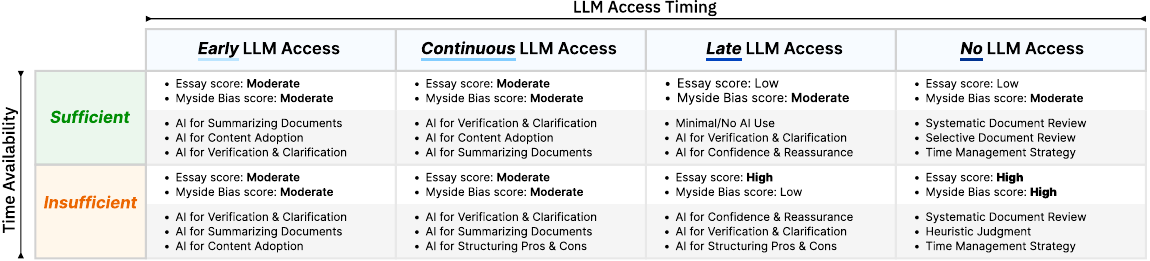}
  \caption{High-level findings across the two-dimensional time constraints framework. White cells show critical thinking task performance as reflected in Essay performance including Essay score and Myside Bias score. Gray cells show common task approaches. Text weight indicates performance level: bold for high, semi-bold for moderate, and regular for low.}
  \label{fig:high_level_findings}
\end{figure*}


\subsubsection{Considering time constraints}

Previous research in thinking and learning examined only portions of the time constraints space in different domains (Table~\ref{tab:related-work-time-constraints} and Figure~\ref{fig:time_constraint_framework}). Our work weaves these fragments together \edit{by viewing them through a unified, two-dimensional lens of time constraints (Figure \ref{fig:high_level_findings}). This perspective extends prior work by considering how their findings might shift under time conditions they did not study. For example, in math education, \citet{kumar2025math} compared early versus late access to LLM explanations under sufficient time, finding that late access improved test performance. What if time availability is insufficient? Under time pressure, these benefits of late access may diminish. It may be hard for users to develop their own understanding first, potentially making late LLM assistance less effective or even disruptive for their learning. In creative writing, \citet{qin2025timing} found that late LLM access preserved originality compared to continuous access under sufficient time. Under time pressure, these benefits may likewise fail to emerge.}



Although participant recruitment requires describing time requirements, task completion is often unconstrained. Even in unconstrained studies, evidence shows that time plays a crucial role in shaping human-AI collaboration. \cite{melumad2025experimental,kreijkes2025llm,jakesch2023co}. 
For example, \citet{jakesch2023co} observed that users spending little time on a writing task with an opinionated in-line AI writing assistant were more influenced by AI opinions.
Moving forward, we suggest that future studies evaluating human-AI collaboration should consider the differential impacts of time constraints. 
Testing under multiple temporal conditions can reveal when benefits emerge or disappear. Our experiment framework (Figure \ref{fig:time_constraint_framework}) provides one approach for systematically considering how combinations of access timing and time availability create distinctions, enabling a more nuanced understanding of when, how, and for whom AI is most beneficial.

Moreover, researchers do not need to examine every possible combination of time constraints; rather, they should select conditions that align with their tasks, design choices, and research goals. In studies where efficiency or time savings is a central outcome, time availability is inherently part of what is being measured \cite{noy2023experimental,bick2025impact,spatharioti2025effects}. \edit{For example, for studies that provide pre-generated LLM responses \cite{kumar2025human}, continuous LLM access is not feasible, but the design can be framed as offering early access and paired with a condition in which participants work independently first. Studies that provide interactive LLMs face additional considerations: for unfamiliar tasks, early and continuous access may effectively converge as users tend to access LLMs from the beginning, though a late-access condition can still meaningfully isolate the value of independent work. In tasks involving participants with diverse expertise, however, users may self-regulate and delay LLM use \cite{dziak2025srt,bo2025s,tankelevitch2025understanding}, meaning continuous access cannot be assumed to function like early access.}

\subsection{Limitations \& Future Work}

While our study provides rich insights on the impact of LLMs on critical thinking, there are several limitations that require further investigation through future work. 

\subsubsection{Task and measurement}

Measuring critical thinking is challenging. We employed a specific task under a well-established performance assessment framework \cite{shavelson2019assessment,braun2020performance} with its arithmetic coding scheme \cite{ebright2024we,EbrightJonesCortina2025}. While validated, there may be limitations when the task is used beyond its original design. Moreover, the civic decision-making scenarios in the framework reflects one context among many. Critical thinking may unfold differently across purposes and domains. For example, debugging code, or making medical diagnoses may draw on distinct reasoning processes. Our task placed participants on equal footing by minimizing the need for specialized background knowledge, but this design choice also limits our ability to generalize to settings where domain expertise substantially shapes performance \cite{dziak2025srt,bo2025s,tankelevitch2025understanding}. Future work should examine domain-specific tasks in which prior knowledge varies to assess whether our findings extend to those contexts.


\subsubsection{Controlled experiment with crowdworkers}

\edit{We conducted our experiment with crowdworkers on Prolific, which allowed for controlled, large-scale data collection but also raises questions about ecological validity. Furthermore, although the performance assessment framework is designed to elicit authentic critical thinking through realistic scenarios \cite{shavelson2019assessment,braun2020performance,EbrightJonesCortina2025}, the study was conducted in a laboratory-style environment with fixed time constraints. Participants did not know their critical thinking performance was being evaluated, yet they were nonetheless aware that they were completing a timed task. In real work contexts, critical thinking unfolds less explicitly: people may not recognize that they are engaging in critical thinking, time constraints arise organically from external or internal pressures, and LLM use is interwoven fluidly throughout the task. Such naturalistic dynamics are challenging to reproduce in controlled experiments. Future work should investigate these processes in settings such as classrooms and workplaces, where tasks carry real stakes and LLM use reflects genuine work practices.}

\subsubsection{Operationalization of time constraints}

Our operationalization of time constraints represents specific design choices within a broader space of possibilities. Alternatives could include graduated time pressure, intermittent LLM access patterns, or user-controlled timing. Future work should explore other temporal configurations.

Our single-session experimental design in a controlled setting helped isolate and disentangle the effects of time constraints, but real-world LLM use is messier than our controlled design could capture. In practice, people may interleave LLM use, or revisit their work at different times. 
Future research should investigate these more fragmented and longitudinal patterns of LLM use.

\section{Conclusion}

This study reveals the fact that LLMs' impact on critical thinking fundamentally depends on time constraints. We found a striking reversal: having LLM access from the start enhanced performance under time pressure but impaired it with sufficient time, while working independently first showed the opposite pattern. Our findings suggest that optimal human-AI collaboration for critical thinking requires careful consideration of when and under what time constraints LLMs should be introduced.
\edit{We hope this work serves as a start for deeper explorations using human-centered experiments to understand AI's impact on critical thinking and human cognition.}


\begin{acks}

We thank our participants for their time and the reviewers for their valuable feedback. 
We also thank Anjali Singh, Jingchao Fang, and Mor Naaman for their insights on the initial versions of the experiment design. We appreciate members of the AI \& Me (AIM) group at the University of Chicago for their helpful feedback.

\end{acks}

\bibliographystyle{ACM-Reference-Format}
\bibliography{main}

\appendix
\section{Appendix}

\subsection{Demographics}\label{demographic_table_appendix}


\begin{table}[htbp]
\centering
\footnotesize
\begin{tabular}{llcc}
\toprule
\textbf{Category} & \textbf{Feature} & \textbf{N} & \textbf{\%} \\
\midrule
\multirow{3}{*}{Gender} 
  & Female & 190 & 48.3 \\
  & Male & 191 & 48.6 \\
  & Non-Binary & 12 & 3.1 \\
\midrule
\multirow{8}{*}{Race/Ethnicity} 
  & White & 265 & 67.4 \\
  & Black or African American & 62 & 15.8 \\
  & Hispanic or Latino & 29 & 7.4 \\
  & Asian & 24 & 6.1 \\
  & Other & 9 & 2.3 \\
  & American Indian or Alaska Native & 2 & 0.5 \\
  & Native Hawaiian or Pacific Islander & 1 & 0.3 \\
  & Prefer not to answer & 1 & 0.3 \\
\midrule
\multirow{6}{*}{Age} 
  & 18–24 & 31 & 7.9 \\
  & 25–34 & 114 & 29.0 \\
  & 35–44 & 113 & 28.8 \\
  & 45–54 & 70 & 17.8 \\
  & 55–64 & 43 & 10.9 \\
  & 65+ & 22 & 5.6 \\
\midrule
\multirow{7}{*}{Education} 
  & Bachelor's degree & 158 & 40.2 \\
  & Master's degree & 69 & 17.6 \\
  & Some college (no degree) & 62 & 15.8 \\
  & Associate degree & 44 & 11.2 \\
  & High school graduate & 43 & 10.9 \\
  & Doctoral degree & 14 & 3.6 \\
  & Professional degree & 3 & 0.8 \\
\bottomrule
\end{tabular}
\caption{Participant demographics}
\label{tab:demographics}
\end{table}

\subsection{Self-Assessment of Critical Thinking}\label{self_assessment_appendix}

\begin{figure}[htbp]
  \centering
  \includegraphics[width=0.45\columnwidth]{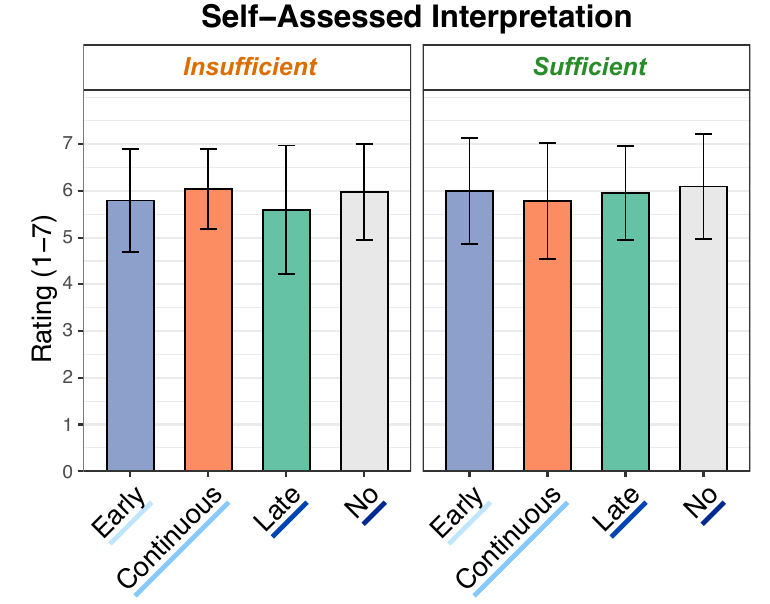}
  \hfill
  \includegraphics[width=0.45\columnwidth]{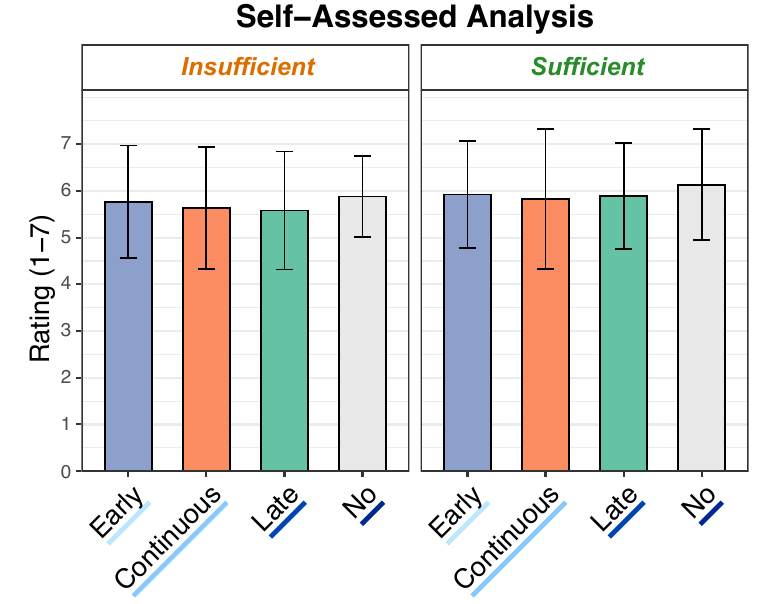}
   
  \includegraphics[width=0.45\columnwidth]{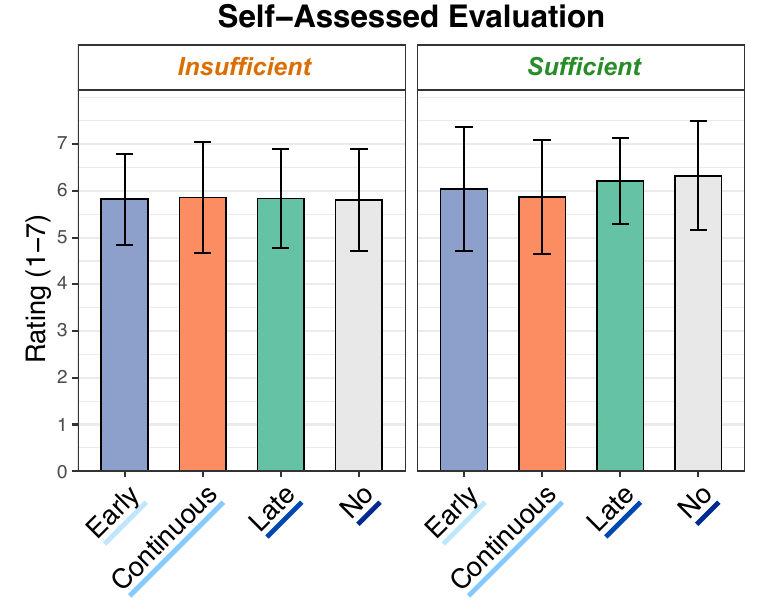}
  \hfill
  \includegraphics[width=0.45\columnwidth]{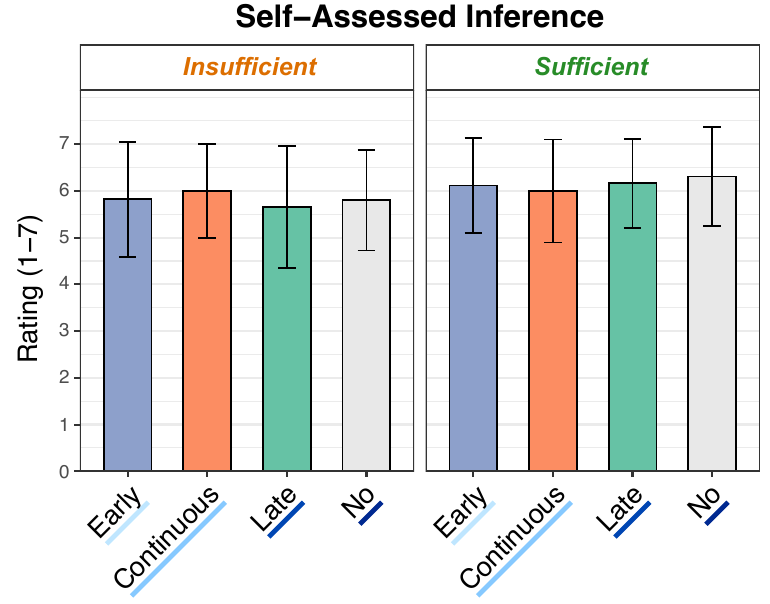}
  
  \includegraphics[width=0.45\columnwidth]{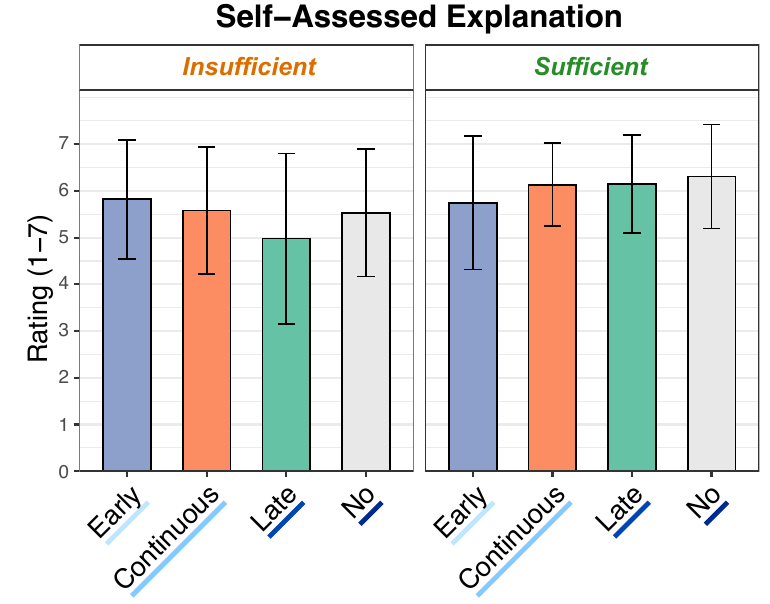}
  \hfill
  \includegraphics[width=0.45\columnwidth]{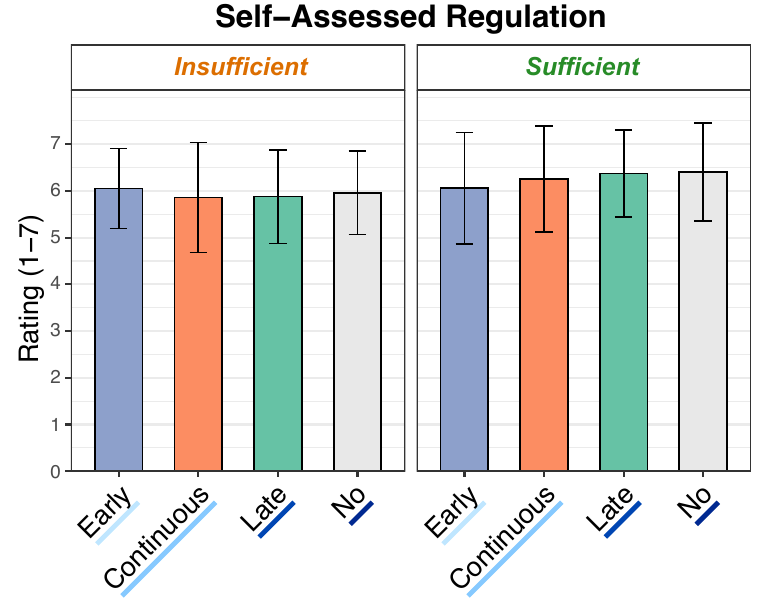}
  
  \includegraphics[width=0.45\columnwidth]{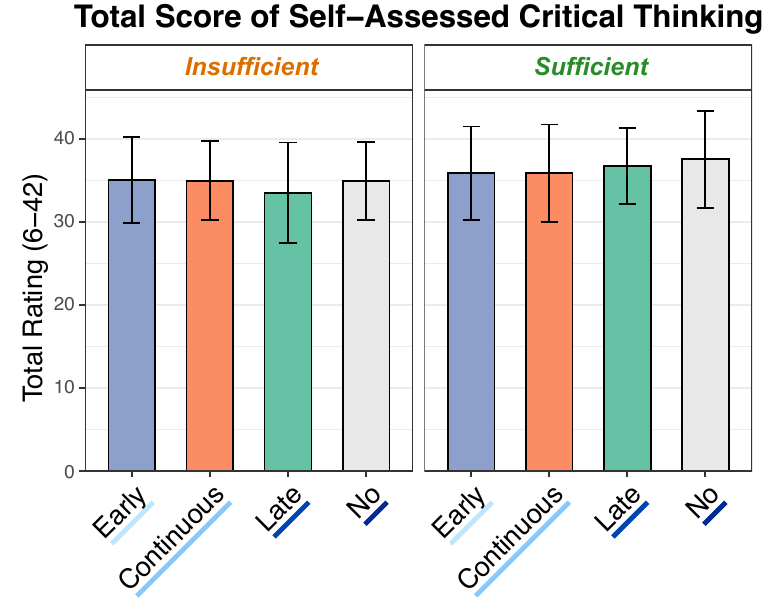}
  \hfill
  \hspace{0.45\columnwidth}

  \caption{Self-assessed critical thinking ratings (mean $\pm$ SD) across conditions.}
\label{fig:self_assessment_plots}
\end{figure}

Figure~\ref{fig:self_assessment_plots} shows participants' self-assessed critical thinking ratings on six components and a total score. Mann-Whitney U tests with Holm-Bonferroni correction revealed significant differences only for time availability: participants under \sufficient time rated higher on interpretation ($U = 11234$, $p.adj = 0.009$), analysis ($U = 11308$, $p.adj = 0.011$), and total score ($U = 11559$, $p.adj = 0.027$). No significant differences were found between LLM access timings.

\subsection{Participants' Task Approaches}\label{description_task_approach}


For participants with LLM access, approaches (1--7) describe how they utilized the LLM. For participants without LLM access, approaches (8--11) describe their task strategies.


\noindent\textbf{1. AI for Confidence \& Reassurance:} Used AI as a check to validate or reassure themselves about arguments or decisions already formed through their own reading and reasoning. 

\noindent\textbf{2. AI for Verification \& Clarification:} Asked AI to explain documents for evidence to support their assumptions and arguments.

\noindent\textbf{3. AI for Summarizing Documents:} Used AI for quick overviews and summaries of documents before or alongside reading.

\noindent\textbf{4. AI for Structuring Pros \& Cons:} Asked AI to organize ideas into pros \& cons or integrate perspectives to weigh trade-offs.

\noindent\textbf{5. AI for Content Adoption:} Directly imported AI-generated text into their essays, such as letting AI draft paragraphs or arguments.

\noindent\textbf{6. AI for Polishing \& Refinement:} Used AI to improve organization, clarity, or wording of their own reasoning essays.

\noindent\textbf{7. Minimal/No AI Use:} Barely used or actively avoided AI, due to uncertainty about how to use it effectively or preference for independent reasoning.

\noindent\textbf{8. Systematic Document Review:} Methodically read through all or nearly all documents in detail, emphasizing thorough coverage.

\noindent\textbf{9. Selective Document Review:} Skimmed across documents or strategically focused only on specific sources deemed most useful.

\noindent\textbf{10. Time Management Strategy:} Employed explicit strategies for dealing with time constraints, such as reading first then writing, prioritizing speed, or adjusting effort allocation.

\noindent\textbf{11. Heuristic Judgments:} Chose documents or formed decisions based on perceived trustworthiness, relevance, bias, or personal intuition rather than comprehensive review.

\end{document}